\documentclass[twocolumn]{openjournal} 

\usepackage{orcidlink} 
\usepackage[T1]{fontenc}
\usepackage{ae,aecompl}
\usepackage[T1]{fontenc}
\usepackage{ae,aecompl}
\usepackage{hyperref}
\usepackage{url}

\usepackage{longtable}
\usepackage{booktabs}
\usepackage{tabu}
\usepackage{graphicx}
\usepackage{multirow}
\usepackage{ulem}
\usepackage{color}
\usepackage{amsmath}

\def \cm{~\rm{cm}}
\def \s{~\rm{s}}
\def \km{~\rm{km}}

\def \g{~\rm{g}}

\def \erg{~\rm{erg}}

\def \mum{~\rm{\mu m}}

\usepackage{xcolor}
\definecolor{redak}{rgb}{0.9,0.15,0.05}
\shorttitle{The jet-shaped pipe morphology}
\shortauthors{Braudo and Soker}

\graphicspath{{./}{figures/}}

\begin{document}

\title{The jet-shaped pipe morphology in planetary nebulae and core-collapse supernova remnants}

\author{Jessica Braudo\,\orcidlink{0009-0001-4877-1125}} 
\affiliation{Department of Physics, Technion - Israel Institute of Technology, Haifa, 3200003, Israel; jessicab@campus.technion.ac.il}

\author{Noam Soker\,\orcidlink{0000-0003-0375-8987}} 
\affiliation{Department of Physics, Technion - Israel Institute of Technology, Haifa, 3200003, Israel; soker@technion.ac.il}

\date{\today}

\begin{abstract}
We compare images of core-collapse supernova (CCSN) remnants (CCSNRs) and jet-shaped planetary nebulae (PNe)  that have a narrow, faint zone extending from side to side, termed a pipe, with a hydrodynamical numerical simulation exploding a massive star with three pairs of jets in the framework of the jittering jets explosion mechanism (JJEM), and conclude that jets shaped the pipes in these CCSNRs and PNe. We present two jet-shaped PNe with a pipe and three PNe with two opposite narrow jet-shaped lobes, and argue that in some cases the two opposite narrow lobes might merge to form one long, faint zone extending from side to side of the PN, namely, a pipe. From the qualitative similarity between the pipe morphology of the two CCSNRs we analyze and the pipe of the PNe, we suggest that jets also shaped the pipe of these CCSNRs.  We strengthen this conclusion with a three-dimensional hydrodynamic simulation that reproduces two opposite narrow lobes, similar to those observed in PNe with lobes. These lobes can merge later to form a pipe. This paper is another in a series that strengthens the case for the JJEM as the primary explosion mechanism of CCSNe by comparing CCSNR morphologies with those of jet-shaped PNe.   
\end{abstract}
   

% ==================================
\section{Introduction} 
\label{sec:intro}
% ==================================

The identifications of similar jet-shaped morphological features in planetary nebulae (PNe) and core-collapse supernova (CCSN) remnants (CCSNRs) in the last decade have strengthened the notion that jets play significant roles in powering and shaping their formation (for reviews see \citealt{Soker2022Rev, Soker2024Rev}). The similar morphological features include barrel-shaped and H-shaped structures (e.g., \citealt{Akashietal2018}), point-symmetric structures, including multipolar structures (e.g., \citealt{BearSoker2018, Soker2022SNR0540, Bearetal2025Puppis}), the presence of opposite pairs of rings (e.g., \citealt{SokerShishkinW49B}), the presence of opposite pairs of ears, rims, and nozzles (e.g., \citealt{BearSoker2017, Bearetal2017, GrichenerSoker2017, Soker2024PNSN, Soker2025G0901}). Point-symmetrical morphologies have two or more pairs of opposite structural features that do not share the same symmetry axis (e.g., \citealt{ShishkinMichaelis2026}). In some cases, the two opposite structural features are not exactly symmetric because, according to the JJEM, jets in a pair during the explosion process might differ substantially in their energy and momentum (e.g., \citealt{Shishkinetal2025S147}). An ear is a protrusion from the main shell, which is smaller than the shell and has a decreasing cross-section with distance. The morphological similarities of hot gas in clusters of galaxies that are known to be shaped by jets with some PNe (e.g., \citealt{SokerBisker2006}) and with some CCSNRs (e.g., \citealt{Soker2024CF}), further solidify the claim that jets shape the above-listed morphological features in PNe and CCSNRs. 

Many of the jet-shaped PNe have close central binary systems (e.g., \citealt{Miszalski2019ic, Jones2020Galax, Jones2025}), indicating they are post-common-envelope evolution (CEE) binaries. One of the main conclusions from jet-shaped PNe with post-CEE binary systems is that jets are the most robust observable ingredient of CEE \citep{Soker2025RobustJetsRAA}, besides the close binary systems that define these systems as post-CEE binaries. This implies that jets play a substantial role in CEE, and that the standard CEE should include, in addition to the orbital energy of the binary systems, the energy of the jets launched by the more compact companion as it enters the envelope, possibly also within the envelope, and as it emerges from the common envelope. Studies show that, in many analyzed post-CEE cases, the orbital energy alone was insufficient to eject the common envelope (e.g., \citealt{Grichener2023, LiZhenweietal2026CE}). In many cases, the jets can supply more energy than the orbital energy released by the binary system.  

The jet-shaped CCSNRs imply that one or more pairs of jets supply most of the energy to the ejecta, as predicted and fully compatible with the theoretical jittering-jets explosion mechanism (JJEM; e.g.,  \citealt{AkashiSoker2026a, Braudoetal2026, Soker2026IJAADust, Soker2026SNRJ0450, Soker2026Failed}, for papers from 2026; for earlier reviews, see, \citealt{Soker2024UnivReview, Soker2025Learning}). On the other hand, the neutrino-driven (delayed neutrino; neutrino heating) explosion mechanism (e.g., \citealt{Akhmetalietal2026, ChenCHetal2026, EggenbergerAndersenetal2026, Giudicietal2026, LuoZhaKajino2026, Mezzacappa2026, Murphyetal2026, PanLi2026, Paradisoetal2026, Rusakovetal2026, VarmaMuller2026, Wessonetal2026}, for several papers from 2026;  \citealt{Janka2025} for an earlier review) cannot account for the morphological properties of point-symmetric CCSNRs. Hydrodynamical simulations of the neutrino-driven mechanism to the CCSNR phase do not address point-symmetric morphologies (e.g., \citealt{OrlandoJankaetal2025A, OrlandoJankaetal2025B, OrlandoMicelietal2025, Orlando2026}). The morphologies of CCSNRs are the only property that can robustly decide between these two heavily studied theoretical explosion mechanisms (e.g., \citealt{Soker2025Learning}). Therefore, the above-listed (and more) papers, which make a solid case that jets shape and power many CCSNRs, have supported the JJEM as the primary explosion mechanism of these CCSNRs, and most likely of all CCSNe;  however, as stated above, the CCSN explosion mechanism is still an open question, with many processes to be explored in the two competing explosion mechanisms. One such process is the shaping of the pipe morphology that we study here.   Our preference for the JJEM is in dispute, as the above-listed papers on the neutrino-driven mechanism present a counter view; the community is far from any consensus on the primary explosion mechanism of CCSNe. 

 Although, as stated, observations at the explosion phase are not at the stage to robustly distinguish between the two explosion mechanisms, there are hints from photospheric evolution. According to the JJEM, in some CCSNe, two or three energetic pairs of jets can form two or three photospheric shells during a CCSN explosion. \cite{SokerShiran2025} argued for two, and possibly three, photospheric shells in SN 2023ixf based on the radius of the photosphere as a function of time as calculated by  \cite{Zimmermanetal2024}, and  \cite{ShiranSoker2026} argued for two photospheric shells, at least one non-spherical, in SN 2024ggi based on the photospheric radius evolution that   \cite{ChenTWetal2025} calculated. These are tentative claims that need further studies and are not as robust as CCSNR morphologies in distinguishing between the two competing explosion mechanisms. 

To further understand the CCSN explosion mechanism by identifying jet-shaped morphological features in CCSNRs, we identify another morphological feature common to some PNe and some CCSNRs, termed a pipe. It is a structure that exists within the inner PN or CCSNR; hence, it cannot be shaped by an interaction with circumstellar material (CSM) or the interstellar medium (ISM). It results solely from the formation process of the PN or the CCSN. Also, it is symmetric around the center and, therefore, it is highly unlikely that instabilities form such a structure. It is most likely a jet-shaped structure. 

The approach we adapt in this study is the same that has been successfully used in understanding the shaping of PNe by jets and binary interaction (e.g.,  \citealt{Morris1987, Soker1990AJ, SahaiTrauger1998, AkashiSoker2018, EstrellaTrujilloetal2019, Tafoyaetal2019, Balicketal2020,   GarciaSeguraetal2020, GarciaSeguraetal2021, Clairmontetal2022, RechyGarciaetal2020, Danehkar2022, MoragaBaezetal2023, Ablimit2024, Derlopaetal2024, Kwok2024Galax, Mirandaetal2024, Sahaietal2024, Masaetal2026}; for a recent review see \citealt{Kwoketal2026Galax}), including precessing jets (e.g., \citealt{Guerreroetal1998, Mirandaetal1998, Sahaietal2005, Boffinetal2012, Sowickaetal2017, RechyGarciaetal2019, Guerreoetal2021, Clairmontetal2022}). The method starts with an eye inspection and classification of PNe based on morphology (e.g., \citealt{Balick1987, Parkeretal2006, Sahaietal2007, Kwok2024Galax}). The qualitative classifications and morphological identifications yield a robust identification of jet-shaped morphological features (e.g., \citealt{SahaiTrauger1998}). 
The jet-like morphological features have driven a wide range of hydrodynamic numerical simulations of PN shaping. The comparisons of simulations with observations (e.g., \citealt{Akashietal2018, GarciaSeguraetal2021, GarciaSeguraetal2022, GarciaSeguraetal2025, Akashietal2025, Kastneretal2025} and references to earlier studies therein) have established jets as a major player in powering and shaping PN outflows. Simulations of the JJEM in the last year showed that jittering jets can reproduce many observed jet-shaped morphological features in CCSNRs \citep{Braudoetal2025, Braudoetal2026, SokerAkashi2025, AkashiSoker2026a, AkashiSoker2026b}.   

%CCSNRs (as well as SN type Ia remnants) are used to infer many aspects of the CCSNRs themselves, e.g., emission properties (e.g., \citealt{LeiXetal2024RAA, Luoetal2024RAA, XianYJetal2025RAA}), interaction of the ejecta with the CSM and ISM (e.g., \citealt{Liangetal2025RAA, Xiaetal2025RAA}) including jets (e.g., \citealt{YuFang2018RAA}), the influence the NS remnant on the CCSNR (e.g., \citealt{HorvathAllen2011RAA, Wuetal2021RAA}), magnetohydrodynamical effects (e.g., \citealt{Wuetal2019RAA, Leietal2024RAA}), dust in CCSNRs (e.g., \citealt{LuTetal2025RAA}), and morphological aspects (e.g., \citealt{Renetal2018RAA}).  We continue our study of CCSNRs to infer the explosion mechanism. 

In this study, we identify a pipe morphological feature that we attribute to a late pair of jets. 
In Section \ref{sec:PipeCCSNRs}, we identify the pipe in CCSNRs, and in Section \ref{sec:PipePNe}, we identify the pipe in PNe. In Section \ref{sec:formingPipes} we present PNe that exhibit morphologies bridging jet-shaped bipolar lobes and pipes, indicating the shaping of pipes by jets.
In Section \ref{sec:Simulation}, we present a preliminary result of a heavy three-dimensional (3D) hydrodynamical simulation under the JJEM framework, which shows the emergence of a pipe. 
We summarize this study in Section \ref{sec:Summary}.

% =========================
\section{Pipes in CCSNRs}
\label{sec:PipeCCSNRs}
% =========================

% =========================
\subsection{The Cygnus Loop}
\label{subsec:CygnusLoop}
% =========================

\begin{figure}[]
\begin{center}
\includegraphics[trim=0cm 13cm 0.0cm 0cm,width=0.62\textwidth]{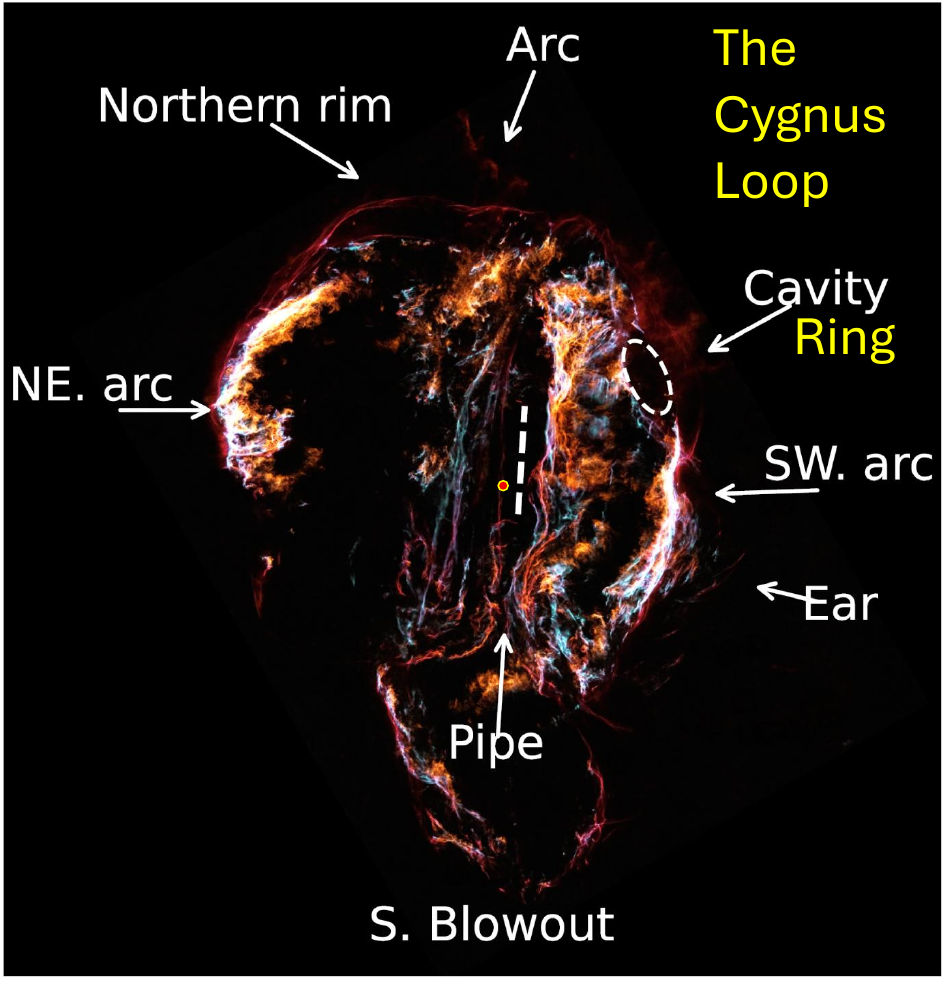} 
\caption{An image of the Cygnus Loop CCSNR in the visible band adapted from \cite{Raymondetal2023}; the marks in white are from \cite{ShishkinKayeSoker2024}, who identified the pipe and a point-symmetric morphology. \cite{SokerAkashi2025} argued that the cavity is a circum-jet ring. The pipe is the long, dark, north-south region; the white-dashed line covers a short portion of it.  The yellow-red dot is the explosion location that \cite{ShishkinKayeSoker2024} estimated.     }
\label{fig:CygnusLoop}
\end{center}
\end{figure}

The motivation to study the pipe morphology stems from the identification of this structure in the Cygnus Loop by \cite{ShishkinKayeSoker2024}, who also identified a point-symmetric morphology with three axes in this CCSNR. 
Figure \ref{fig:CygnusLoop} is a visible image of the Cygnus Loop CCSNR adapted from \cite{Raymondetal2023}. \cite{ShishkinKayeSoker2024} identified the pipe as the dark, narrow north-south region between bright filaments. They claim it was shaped by a pair of opposite jets. It is also visible in the UV (not shown here). At its ends, the pipe bent to the west in the north and to the east in the south, forming a large-scale `S-shaped' structure. 
% FFFFFFFFFFFFFFFFFFFFFFFFFFFFFFFFFFFFFFFFFF
% FFFFFFFFFFFFFFFFFFFFFFFFFFFFFFFFFFFFFFFFFF

  There is no identification of a central star remnant in the Cygnus Loop. \cite{ShishkinKayeSoker2024} uses the centers of the three symmetry axes to estimate the location of the explosion to be within the pipe, slightly displaced east of the center of the pipe, as we mark with a yellow-red dot in Figure \ref{fig:CygnusLoop}. In that respect, the pipe is compatible with two opposing jets from the newly born neutron star (NS) in the Cygnus Loop explosion. However, without kinematic data, our identification of the pipe is still tentative. We here will follow \cite{ShishkinKayeSoker2024} in referring to the pipe as a jet-shaped structure.   

The importance of the pipe is that it exists inside the nebula, and therefore, the interaction of the CCSN ejecta with the CSM or the ISM cannot shape it. It must result from the explosion process. This structure is compatible with the JJEM. 
An interaction of the ejecta with a previously shaped CSM might lead to the formation of a structure like the blowout to the south (e.g., \citealt{ShenJYetal2024RAA}). However, \cite{ShishkinKayeSoker2024} pointed out that the pipe connects to the south blowout, and there might also be one in the north. If the pipe is related to the blowout, then the blowout was also a result of the explosion, i.e., inflated by a jet as  \cite{ShishkinKayeSoker2024} suggested. 

The goal of this study is to highlight similarities with pipes in PNe, thereby solidifying the claim that the pipe morphology results from a pair of opposite jets. The boundaries of the pipe of the Cygnus Loop are not straight, but rather, they wiggle. Several processes can cause wiggling, including the effects of other jets in the explosion, instabilities that occur as the shock expands through the star, and nickel bubbles (local heating from the radioactive decay of nickel clumps).  
  
% =========================
\subsection{SNR G292.0+1.8}
\label{subsec:G292}
\begin{figure}[]
\begin{center}
\includegraphics[trim=3.7cm 0cm 0.0cm 0cm,width=0.55\textwidth]{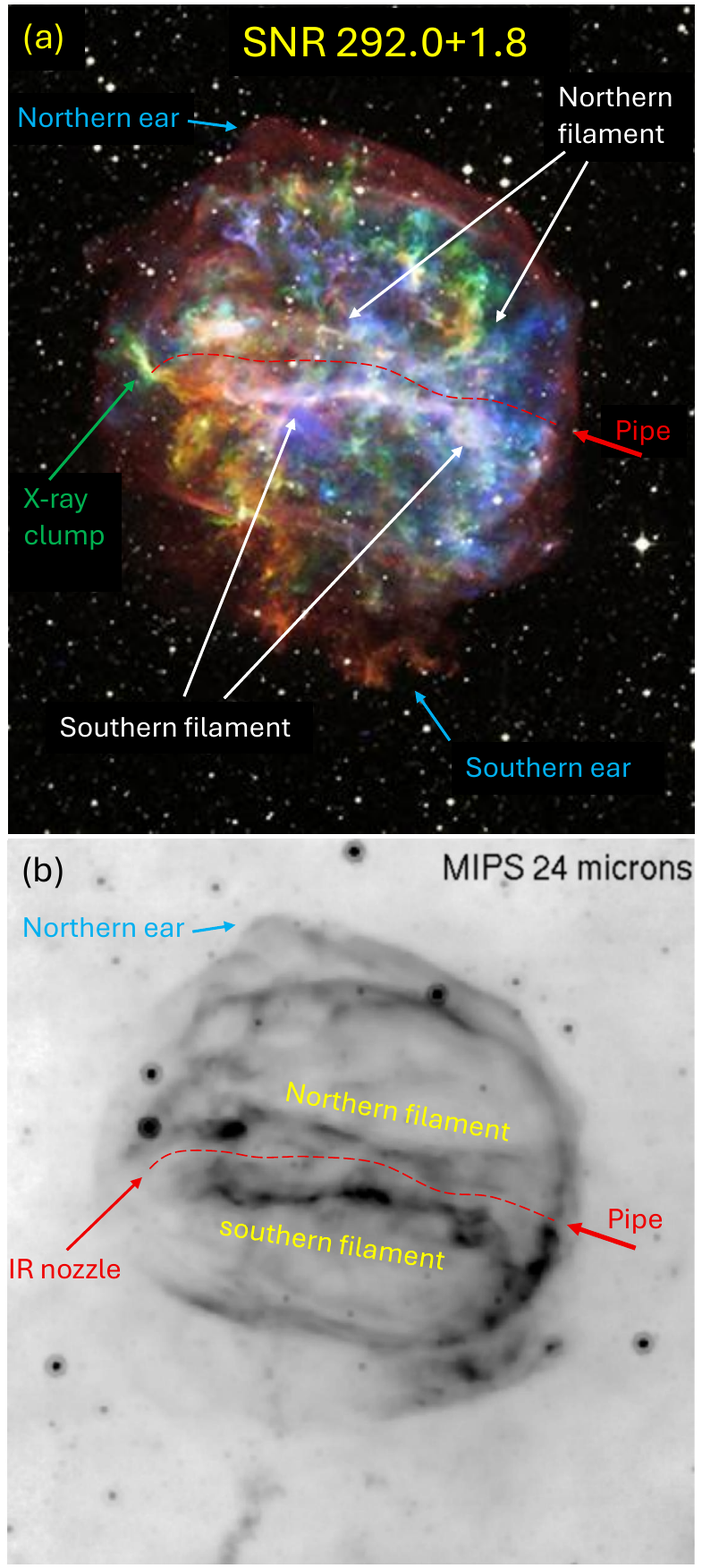} 
\caption{Images of SNR G292.0+1.8. (a)  An image from the Chandra site (Credit: X-ray: NASA/CXC/Penn State/\citealt{Parketal2007}; Optical: Pal.Obs. DSS): Red (0.580-710 and 0.880-950 keV), Orange (0.980-1.100 keV), Green (1.280-1.430 keV), Blue (1.810-2.050 and 2.400-2.620 keV); Optical (white). 
We added labeling for the two ears identified by \cite{Bearetal2017} as jet-shaped structures, as well as for the north and south filaments we study here. We suggest that the filaments here and in the PNe we study are the limb-brightened projection of a cylindrical surface around the pipe. 
The pipe that we identify here is the long east-west region between the filament; the red-dashed line is copied from panel (b). 
(b) An IR \textit{Spitzer} image in $24 \mum$ adapted from \cite{Ghavamianetal2012}; we added the labeling. The dashed red line marks the pipe. 
}
\label{fig:G292}
\end{center}
\end{figure}

% =========================
SNR G292.0+1.8 is a well-studied CCSNR (e.g., \citealt{Temimetal2022, Naritaetal2024, Plunkettetal2026} and earlier references therein; for its pulsar, see, e.g., \citealt{LonkXietal2022, Lemiereetal2026}). 
Relevant to the JJEM is the identification of a pair of ears by \cite{Bearetal2017}, who attributed the morphology to a pair of jets. The line connecting the two ears and the line through the pipe that we identify here, form a point-symmetric morphology, which, according to the JJEM was shaped by at least two pairs of jets that participated in the explosion process. We postpone the study of the full point-symmetric morphology of SNR G292.0+1.8 to a future study, and in the present study, focus on the two bright filaments extending from the east to the west sides of the main CCSNR shell.
We present images of SNR G292.0+1.8 in Figure \ref{fig:G292}, and point at the northern and southern filaments, and mark the pipe that we identify.

% FFFFFFFFFFFFFFFFFFFFFFFFFFFFFFFFFFFFFFFFFF
% FFFFFFFFFFFFFFFFFFFFFFFFFFFFFFFFFFFFFFFFFF

Several studies identified these two filaments as a ring and termed it the equatorial ring or equatorial belt (e.g., \citealt{Parketal2002, Parketal2004, Ghavamianetal2005, Ghavamianetal2012, Leeetal2009}).  These studies attributed it to a CSM because it is not metal-rich. 
However, \cite{Bhaleraoetal2019} noted that the two filaments do not have the limb-brightening expected for a pure ring structure. Namely, the east and west ends, where the filaments are supposed to meet, are not brighter than the rest of the filaments. 
\cite{Bhaleraoetal2019} studied the properties of the filaments, which they also refer to as the equatorial ring and consider to be a CSM lost by the CCSNR progenitor in the equatorial plane; they estimated its mass as $\simeq 1.7 M_\odot$. 
The problem with interpreting the two filaments as a ring is more evident in IR images. These IR images (e.g., \citealt{Leeetal2009, Ghavamianetal2012}), one of which we present in panel (b) of Figure \ref{fig:G292}, show that the two filaments do not meet at their ends as expected for a ring if one filament is behind and one in front of the main SNR shell, not only in the west, but also in the east; we mark this open end as an IR nozzle. We drew a thin red dashed line through the approximate center of the pipe, as shown in the IR image (panel b), and copied it onto the X-ray image (panel a). The wiggling of the pipe is qualitatively similar to the boundary of the pipe in the Cygnus Loop. 

The X-ray map is not identical to the IR map. For example, there is a bright X-ray clump at the east end of the pipe where the IR nozzle is (it is very clearly seen in figure 4 of \citealt{Leeetal2010}). The clump has a different color from the filaments; it is green (with higher X-ray energies than the filaments). \cite{GonzalezSafiHarb2003} find that this clump has higher ionization state values than the filaments. 
We suggest this clump is related to the pipe, but it is not a continuation of the filaments.

We attribute the pipe to the narrow columns formed by two opposite jets as they propagate outward from the center. The two opposite jets compressed the stellar material, mainly from the envelope, to form the filaments (which envelope the pipe). 
\cite{Naritaetal2024} find that the filaments have a nitrogen-to-oxygen abundance ratio lower than solar and argue that this indicates heavy pre-explosion mass loss; they argued the explosion was a type Ib/c CCSN. 
We attribute the filaments to a still intact stellar envelope that the jets compressed. 

 Our proposed identification of a pipe in SNR G292.0+1.8 enclosed by the northern and southern filaments is tentative. As in the case of the Cygnus Loop, without kinematic data, this identification cannot be considered robust.  

The filaments are also bright at $70 \mum$ (e.g., \citealt{Ghavamianetal2012}), indicating dust emission. In the scenario we propose, the jets that shaped the pipe compressed stellar material during the explosion, thereby promoting dust formation. Other CCSNRs indicate that jittering jets in CCSNe promote dust formation \citep{Soker2026IJAADust}. 

Our claimed jet axis is the third one proposed for SNR G292.0+1.8. \cite{Bearetal2017} identified a jet axis that connects the two ears; it points from south-southwest to north-northeast. \cite{Plunkettetal2026}, who conduct an optical Doppler shift study, find a bipolar outflow somewhat inclined to the first axis, pointing from south-southeast to north-northwest. We now suggest a third axis, due to late jets. Overall, the morphology of SNR G292.0+1.8 is multipolar (point-symmetric), not bipolar, as the JJEM predicts for many CCSNRs. The study of the point-symmetric structure of SNR G292.0+1.8 is the subject of a forthcoming paper.

% =========================
\section{Pipes in PNe}
\label{sec:PipePNe}
% =========================

The Ring Nebula (NGC 6720) is an iconic PN with several tens of studies of its morphology over the years (e.g., \citealt{Clarketal2025, Kastneretal2025Ring, Sahaietal2025, Wessonetal2024} for some more recent studies). We here focus only on its pipe morphology. 

\begin{figure*}[]
\begin{center}
\includegraphics[trim=0.0cm 15.70cm 0.0cm 0.0cm,width=1.00\textwidth]{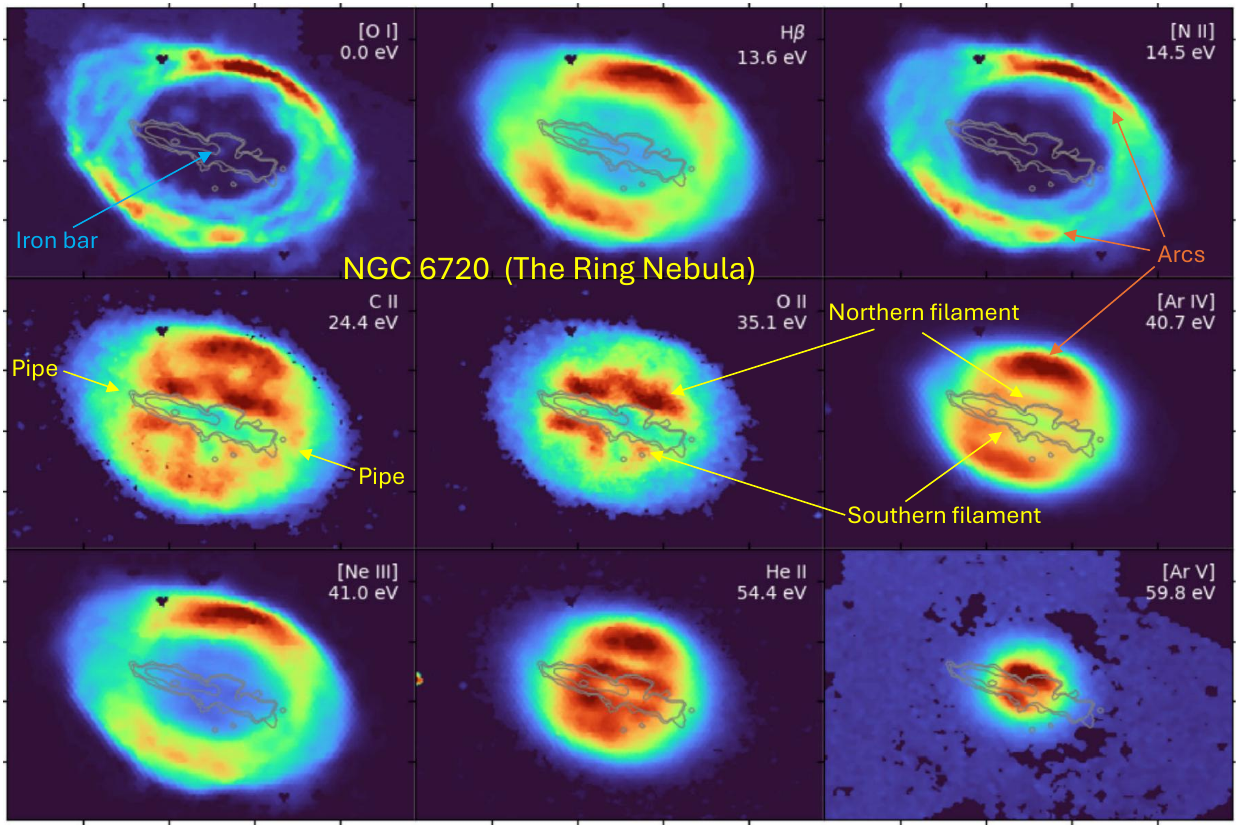} 
\caption{Emission-line maps of PN NGC 6720 adapted from \cite{Wessonetal2026TheRingPN}, who conducted and analyzed observations by the William Herschel Telescope on La Palma. 
Each map indicates the emitting species and the associated creation energy. The contours are of [Fe \textsc{v}] emission and represent the iron bar; collisional excitation of Fe$^{4+}$ (54.8 eV ionization potential) produces this line. 
Images are on a linear surface brightness scale.
We added the marks of the pipe, filaments, and arcs. We consider the filaments to be the limb-brightened projection of a cylindrical thin surface bounding the pipe. The vertical axis is the declination with four marks from $33^\circ 01^{\prime}20^{\prime \prime}$ to $33^\circ 02^{\prime}20^{\prime \prime}$, and the horizontal axis is the right ascension with four marks from 
$18^{\rm h}53^{\rm m}38^{\rm s}$ to $18^{\rm h}53^{\rm m}32^{\rm s}$
}
\label{fig:NGC6720Vis}
\end{center}
\end{figure*}

\begin{figure}[]
\begin{center}
\includegraphics[trim=0.0cm 13.80cm 0.0cm 0.0cm,width=0.94\textwidth]{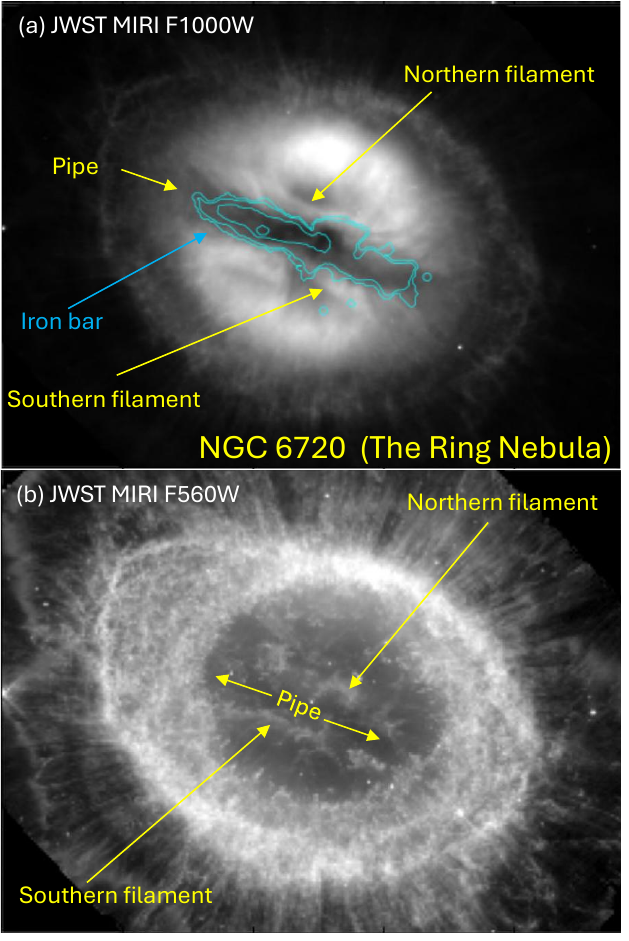} 
\caption{
IR JWST images of NGC 6720 adapted from  \cite{Wessonetal2026TheRingPN} and based on \cite{Wessonetal2024}. We added the marks of the filaments and pipe. (a) MIRI F1000W with contours of [Fe \textsc{v}] 4227 \AA. (b) MIRI F560W, which emphasizes H$_2$ emission. 
}
\label{fig:NGC6720IR}
\end{center}
\end{figure}

In a recent study by \cite{Wessonetal2026TheRingPN}, detailed images of NGC 6720 were presented, revealing a narrow east-west bar of [Fe \textsc{v}] and [Fe \textsc{vi}] emission. This bar sits just between two filaments, or just inside what we term here pipe morphology. We suggest that here, as in the SNRs we study here and the rest of the PNe, the filaments are the limb-brightened projection of a thin cylindrical surface bounding the pipe. We present some images in Figures \ref{fig:NGC6720Vis} and \ref{fig:NGC6720IR}, in the visible and IR, respectively. We added the marks of the pipe and of the northern and southern filaments, which are the boundaries of the pipe.  These two figures reveal the following properties, similar to those of the pipes in CCSNRs as presented in Figures \ref{fig:CygnusLoop} and \ref{fig:G292}.  (1) The projected boundaries of the pipe are not straight, but rather wiggly. (2) The pipe is very prominent in some wavelengths and bands, while almost unseen in others. (3) There is an opening on each of the two ends of the pipe. 
% FFFFFFFFFFFFFFFFFFFFFFFFFFFFFFFFFFFFFFFFFF
% FFFFFFFFFFFFFFFFFFFFFFFFFFFFFFFFFFFFFFFFFF
% FFFFFFFFFFFFFFFFFFFFFFFFFFFFFFFFFFFFFFFFFF
% FFFFFFFFFFFFFFFFFFFFFFFFFFFFFFFFFFFFFFFFFF

\cite{Wessonetal2026TheRingPN} who discovered and studied the iron bar, considered its structure to be a puzzle for its highly ionized state, and for its velocity. They find that the iron bar is redshifted relative to its surroundings. Although \cite{Wessonetal2026TheRingPN} find no bipolar velocity structure, \cite{Kastneretal2025Ring} show velocities in CO that reveal that the east side of the northern filament is blueshifted by $-54$ to $-36 \km \s^{-1}$, while the west side of the southern filament is redshifted by $+34$ to $+52 \km \s^{-1}$. This is a bipolar velocity structure. 
We suggest that two opposite jets shaped and inflated the pipe (Section \ref{sec:formingPipes}), and that the iron bar is a highly ionized gas that fills the pipe. Its origin is the hot central star, which explains its highly ionized state. A high iron abundance might result from the destruction of Earth-like planets or smaller solid bodies by the central star. 

\begin{figure}[]
\begin{center}
\includegraphics[trim=0.0cm 22.8cm 0.0cm 0.0cm,width=0.94\textwidth]{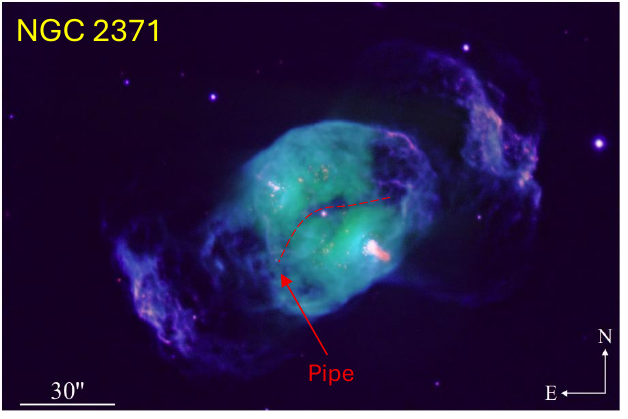} 
\caption{
An image of the multipolar PN NGC 2371 adapted from \cite{GomezGonzalezetal2020}. We added the dashed red line to indicate the pipe. Red, green, and blue correspond to [N \textsc{ii}], H$\alpha$, and [O \textsc{iii}],
 respectively.
}
\label{fig:NGC2371}
\end{center}
\end{figure}

Another PN with a pipe morphology is NGC 2371, which we present in Figure \ref{fig:NGC2371} adapted from \cite{GomezGonzalezetal2020}, who studied its structure in detail and identified it as a multipolar PN. In a more recent paper, \cite{Vazquezetal2026} thoroughly study its structure and also consider it to be a multipolar PN. 
We focus only on the pipe that we define here, and emphasize that, in this case, the pipe is not straight but bent in the inner region of the PN. There are no clear filaments on the boundary of the pipe. There are two low-ionization opposite knots (red) at the main PN shell. They define an axis through the center (southwest to northeast), which is inclined to the pipe. \cite{GomezGonzalezetal2020} argued that their slow motion and structure show they are not jets. Their presence suggests that an observer with a line of sight through the knots would not be able to identify the pipe. Namely, in many cases, a pipe exists in a PN or a CCSNR, but we cannot identify it.
There are other PNe similar to NGC 2371, but not identical, e.g., NGC 6804 (images in, e.g., \citealt{Schwarzetal1992}). 
% FFFFFFFFFFFFFFFFFFFFFFFFFFFFFFFFFFFFFFFFFF
% FFFFFFFFFFFFFFFFFFFFFFFFFFFFFFFFFFFFFFFFFF

Overall, we argue that the pipe of PN NGC 6720 has significant morphological similarities with the pipe of the Cygnus Loop, and the pipe of NGC 2371 with that of SNR G292.0+1.8. We show next that jets shape the pipes in PNe, implying that they most likely also shape the pipes in CCSNRs, as predicted by the JJEM. 

% =========================
\section{Forming a pipe with a pair of jets}
\label{sec:formingPipes}
% =========================

To demonstrate from observations that jets can shape the pipe morphology, we present PNe and pre-PNe with a structure consisting of two lobes that form a pipe, or part of a pipe. 

\begin{figure}[t]
\begin{center}
\includegraphics[trim=0.0cm 20.3cm 0.0cm 0.0cm,width=0.94\textwidth]{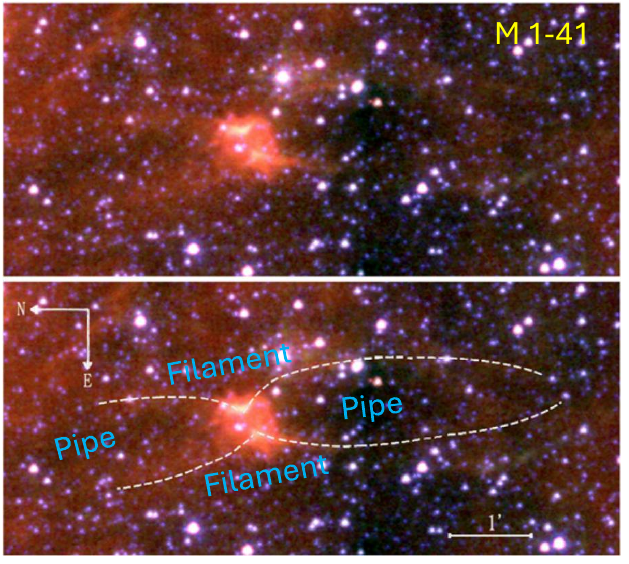} 
\caption{An IR Spitzer image of PN M 1-41 adapted from \cite{ZhangYetal2012}: blue, green and red correspond to $3.6 \mum$, $5.8 \mum$, and $8.0 \mum$, respectively. They drew the dashed lines to sketch the extended bipolar structures. We term them filaments and term the pipe. 
}
\label{fig:M141}
\end{center}
\end{figure}

Figure \ref{fig:M141} that we adapted from \cite{ZhangYetal2012} shows an IR image of the PN M1-41. This is a bipolar PN, most likely formed by jets. The lower panel is the same as the upper one, with the dashed lines drawn by \cite{ZhangYetal2012} to emphasize the extended bipolar structure. These lines are similar to the filaments that bound the pipe in the PN NGC 6720 and in the CCSNRs, the Cygnus Loop, and SNR G292.0+1.8. Here, the pipe is clearly narrower near the center. At much later times, as the entire nebula expands, the difference between the narrow zone near the center and the rest of the pipe will diminish. 
This morphology type connects the classical two-lobes structure with a very narrow waist between them and the pipe morphology. It suggests that pipes are the merger of two long jet-inflated lobes.  
% FFFFFFFFFFFFFFFFFFFFFFFFFFFFFFFFFFFFFFFFFF
% FFFFFFFFFFFFFFFFFFFFFFFFFFFFFFFFFFFFFFFFFF

\begin{figure}[]
\begin{center}
\includegraphics[trim=0.0cm 23.3cm 0.0cm 0.0cm,width=0.94\textwidth]{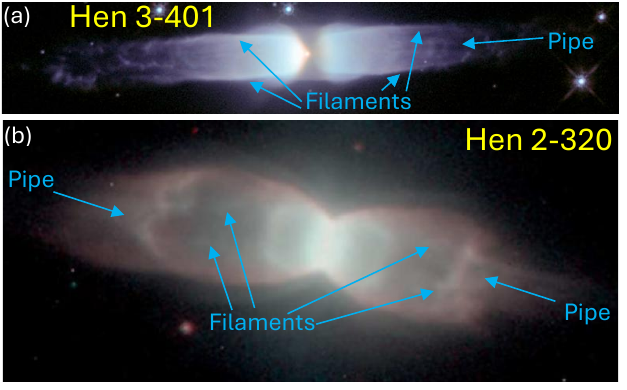} 
\caption{
Planetary nebulae that show two jet-shaped lobes (or more) that we suggest might later evolve to one long pipe. 
(a) An HST image of the young PN Hen 3-401 (Credit: European Space Agency and Pedro Garcma-Lario at ESA ISO Data Center)  
(b) A composite image of PN Hen 2-320 adapted from \cite{Hsiaetal2014}: blue is H$\alpha$, green is H$\alpha$+[N \textsc{ii}], and red is [N \textsc{ii}]. We added the marks of filaments and the pipe. 
}
\label{fig:SeveralPNe}
\end{center}
\end{figure}

There are several pre-PNe and very young PNe that show two elongated bipolar lobes that, at later times, might merge into one pipe, or when viewed not perpendicular to their axis, might appear as one long pipe, e.g., Hen 3-401. The H$_2$  images of Hen 3-401 that \cite{Hrivnaketal2008} present show such a structure, with two bright lobes, each with two clear filaments that bound a pipe. We here present an \href{https://esahubble.org/images/heic0209a/#:~:text=This%20image%2C%20taken%20with%20the%20NASA/ESA%20Hubble,extraordinary%20vision%20reveals%20that%20it%20is%20one}{HST image of Hen 3-401} in panel (a) of Figure \ref{fig:SeveralPNe}. We mark the filaments and pipes, although at this early stage of evolution, the two lobes have separate pipes along the same axis. In the future, they might be merged into a single long pipe.  There are other similar PNe (e.g., \href{https://esahubble.org/images/potw1606a/}{Hen 2-437}).   
% FFFFFFFFFFFFFFFFFFFFFFFFFFFFFFFFFFFFFFFFFF
% FFFFFFFFFFFFFFFFFFFFFFFFFFFFFFFFFFFFFFFFFF

The PN Hen 2-320 is a multipolar PN, with possible four bipolar pairs more or less along the same axis \citep{Hsiaetal2014}. The multipolar structure strongly indicates shaping by pairs of jets. In panel (b) of Figure \ref{fig:SeveralPNe}, we present an image of this PN adapted from \cite{Hsiaetal2014}. There are limb-brightened lobes, namely, two filaments on the boundary of each lobe. 
At later evolutionary times, the central bright ring might disappear, leading to the merging of the two opposite lobes into two long filaments with a long pipe between them. 
The PN M 2-46 (not shown here) is another PN with a pair of lobes within a pair of lobes, again, indicating shaping by two pairs of jets. This structure might form a pipe, bound by two filaments, within the outer lobes.  

For the pre-PN M 1-92 (not shown here), a structure resembling a pipe is deduced from structure reconstruction from velocity measurements. Its visible image (e.g., \citealt{LiYQMorrisSahai2024Galax}) shows two prominent lobes in a bipolar structure that suggests shaping by jets. Reconstruction of its structure by the velocity along the line of sight shows elongated lobes with a dense equatorial material between them (e.g., \citealt{Alcoleaetal2022Galax, Masaetal2024Galax}). At a much later time, if the equatorial material expands, the two lobes will merge to form a long pipe. 

The PNe we discussed in this section, and others, have morphologies of two narrow, jet-shaped lobes. If the dense material in the equatorial plane expands and its emission decreases, the two lobes might merge, appearing as a pipe. We take these to suggest that the pipe morphology is the merger of two long, narrow lobes.    Again, without detailed kinematical data, our suggestion is still speculative. We encourage more observations to reveal and explore the pipe morphology properties. 

% ==================================
\section{Simulating pipe formation in the JJEM} 
\label{sec:Simulation}
% ==================================
% ==================================
\subsection{The numerical scheme} 
\label{subsec:NumericalScheme}
% ==================================
We present preliminary results of a simulation that shows the emergence of a pipe in the JJEM. This simulation from \cite{Braudoetal2026} was not intended to form a pipe. The pipe emerges in the simulation, depending on the viewing angle. The study of pipe formation is a heavy task that we leave for a future study. 
We present here only the essential numerical information  (for more details, see \citealt{Braudoetal2026}) and the preliminary results showing the emergence of a pipe.

This simulation (termed E3 in \citealt{Braudoetal2026}), within the framework of the JJEM, includes three pairs of equal and opposite jets. The simulation was performed with version 4.8 of the {\sc flash} hydrodynamical code \citep{Fryxell2000} in 3D. \cite{Braudoetal2026} launch the three jet pairs into an initially spherically symmetric Wolf–Rayet (WR) stellar model of a collapsing massive stellar core. The innermost region of the initial stellar model $r \le 2\times10^{9} \cm$ ($20,000 \km$), is a collapsing core from \citet{PapishSoker2014Planar}, who based this model on a $t \simeq 0.2 \s$ post-bounce structure of a $15\,M_\odot$ progenitor \citep{Liebendorferetal2005}.  
The initial region of $2\times10^{9} \cm \le r \le 8\times10^{9} \cm$ is an envelope of a hydrogen- and helium-stripped $15\,M_\odot$ WR progenitor model. The density at the stellar surface is $\simeq 30 \g \cm^{-3}$.
Outside the stellar model, $r > 8\times10^{9} \cm$, the numerical CSM density decreases. In this study, we are interested in the dense inner parts of the ejecta, so the CSM is less relevant. The simulation includes the gravitational field of a newly formed $1.4\,M_\odot$ point-mass NS.
The launching radius of the jets is $5\times10^{7} \cm$. 

Each of the six jets carries the same mass and energy, ${m}_{\rm j}=1.6 \times 10^{31} \g$ and $E_{\rm k,j} =  2 \times 10^{50} \erg$, respectively, and is active for the same duration $\Delta t_{\rm j}({\rm active}) = 0.08 \s$. The total kinetic energy injected by all six jets is $E_{\rm k,tot} =  1.2 \times 10^{51} \erg$. The expected initial velocity of the jets in the JJEM is $v_{\rm j,JJEM} \simeq 10^5 \km \s^{-1} - 1.5 \times 10^5 \km \s^{-1}$. However, to reduce computational time (which is already very long), we inject the jets only at $v_{\rm j} = 5 \times 10^4 \km \s^{-1}$.
To maintain the explosion energy, we had to increase the mass that each jet carries. For numerical reasons, we also inject the jets at a radius about an order of magnitude larger than the real launching radius.   
Table \ref{Tab:Table} lists the time and direction of each jet; for more details, see \cite{Braudoetal2026}. 
The last pair of jets is launched along the $y$-axis, i.e., along $(x,z)=(0,0)$, and it will shape the emerging pipe. 
% TTTTTTTTTTTTTTTTTTTTTTTTTTTTTTTTTTTTTTTTTTTTTTTTTTTTT
\begin{table}[h]
%\tiny
\scriptsize
%\footnotesize
\begin{center}
  \caption{Jet parameters}
\begin{tabular}{|p{2.0cm}|c c|c c|c c|}
\hline
Jets' pair & \multicolumn{2}{c|}{1} & \multicolumn{2}{c|}{2} & \multicolumn{2}{c|}{3} \\
\hline
Jet         & + & - & + & - & + & - \\
\hline
$t_{\rm j}({\rm start})\ [\rm s]$ & $0.00$ & $0.00$ & $0.48$ & $0.48$ & $0.96$ & $0.96$\\
\hline
$\theta_{\rm j}\ [\rm deg]$ & $35$ & $145$ & $0$ & $180$ & $90$ & $90$ \\
\hline
$\phi_{\rm j}\ [\rm deg]$ & $0$ & $180$ & $0$ & $0$ & $90$ & $270$ \\
\hline
\end{tabular}
  \\
\label{Tab:Table}
\end{center}
\begin{flushleft}
\small 

Notes: The properties of the six conical jets launched in three equal-jet pairs. All jets are launched at a velocity of $v_{\rm j} = 50,000 \km \s^{-1}$, which amounts to a Mach number of $\simeq 6.5$, have an activity duration of $\Delta t_{\rm j}({\rm active}) = 0.08 \s$, and a half opening angle of $\alpha_{\rm j}= 10^\circ$. 
The mass injection rate into each jet is $\dot{m}_{\rm j}=2 \times 10^{32} \g \s^{-1}$, and the kinetic energy it carries is $E_{\rm k,j} =  2 \times 10^{50}  \erg$. The total kinetic energy injected by all six jets is $E_{\rm k,tot} =  1.2 \times 10^{51} \erg$.
The first and second rows give the name of the pair and jet, respectively. 
The following rows list the starting time $t_{\rm j}({\rm start})$ and the direction of each jet $(\theta_{\rm j}, \phi_{\rm j})$, where $\theta_{\rm j}$ is the polar angle measured from the $z$-axis, and $\phi_{\rm j}$ is the azimuthal angle measured in the $xy$-plane relative to the $x$-axis.  
\end{flushleft}
\end{table}
% TTTTTTTTTTTTTTTTTTTTTTTTTTTTTTTTTTTTTTTTTTTTTTTTTTTTT

% ==================================
\subsection{The emergence of a pipe} 
\label{subsec:TheemergencePipe}
% ==================================

Since we do not follow the pipe for a long time (due to numerical limitations), we refer to the pipe identification in the 
simulation as the emergence of a pipe. We present the emergence of the pipe from a single viewing angle. In a future paper, we will present other viewing angles, other jets' parameters, and follow the evolution to much later times.

\begin{figure}[]
\begin{center}
\includegraphics[trim=1.0cm 0.5cm 0.6cm 0.0cm ,clip, angle=0, scale=0.9]{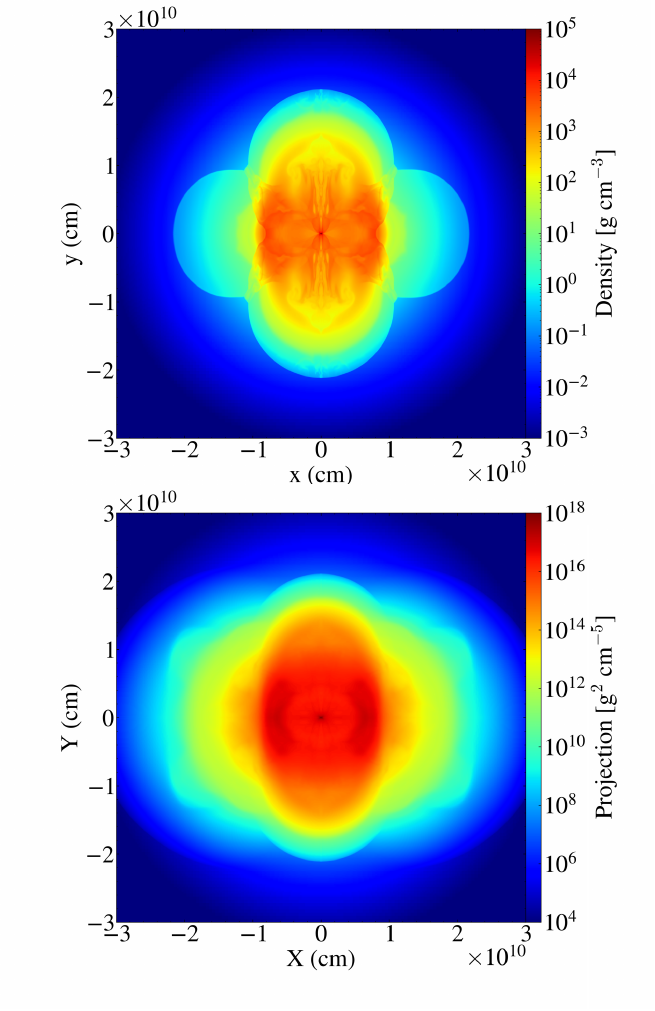} 
\caption{Results from the 3D hydrodynamical simulation. Upper panel: Density map in the $z=0$ plane according to the color bar in logarithmic scale from $10^{-3} \g \cm^{-3}$ to $10^5 \g \cm^{-3}$ (deep red). The third pair of jets is launched in this plane along the $y$-axis. 
Lower panel: The emission integral, which is the synthetic observation map, along the the $z$-axis (equation \ref{eq:proj}); values according to the color bar in logarithmic scale from $10^{4} \g^2 \cm^{-5}$ (deep blue) to $10^{18} \g^2 \cm^{-5}$ (deep red). Both images are at the end of the simulation, $t=13 \s$. 
}
\label{fig:SimulationDensproj}
\end{center}
\end{figure}

In the upper panel of Figure \ref{fig:SimulationDensproj}, we present the density map on the plane $z=0$ at $t=13 \s$; colors indicate density on a logarithmic scale. The last pair of jets lies along the $y$-axis (vertical in the plane of the upper panel); one jet starts in the $+y$ direction, and the other in the $-y$ direction. The density map shows dense filaments near the $y$-axis, but not very close to it. The low-density region along the $y$-axis is the pipe. To mimic observations, we take the emissivity (power per unit volume) to be proportional to the square of the density, $\epsilon \propto \rho^2$. The numerical (scaled) intensity on the plane of the sky for an observer along $L$ is the numerical emission integral 
\begin{equation}
    {\rm EI}(X,Y) = \int \rho^2(x,y,z) \, d L,
    \label{eq:proj}
\end{equation}
where $\rho(x,y,z)$ is the density, $(X,Y)$ are the coordinates on the plane of the sky perpendicular to the line of sight, and the integration is along the line of sight $L$ that in this study is along the $z$ direction. 
We present the numerical emission integral in the lower panel of Figure \ref{fig:SimulationDensproj}. This is the synthetic observation image. 
The emission integral shows two arcs to the sides of the $y$-axis (black arcs inside the red region). This type of structure is commonly seen in jet-shaped planetary nebulae, e.g., the upper row of Figure \ref{fig:NGC6720Vis}.   
% FFFFFFFFFFFFFFFFFFFFFFFFFFFFFFFFFFFFFFFFFF
% FFFFFFFFFFFFFFFFFFFFFFFFFFFFFFFFFFFFFFFFFF

\begin{figure}[]
\begin{center}
\includegraphics[trim=0.6cm 0.4cm 0.3cm 0.0cm ,clip, angle=0, scale=1.3]{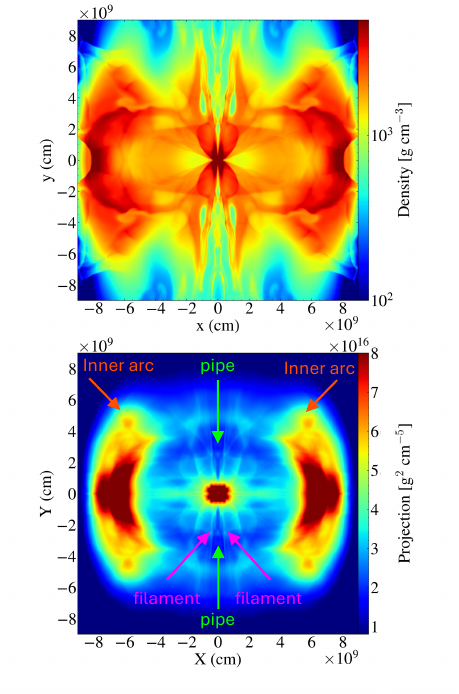} 
\caption{Similar to Figure \ref{fig:SimulationDensproj} but focusing on the inner region, and using a linear scale for the emission integral in the lower panel. The density logarithmic scale, from deep blue to deep red, is $10^{2} - 5 \times10^3 \g \cm^{-3}$. The emission integral linear scale, from deep blue to deep red, is $8 \times 10^{5} - 8 \times 10^{16} \g^2 \cm^{-5}$. The high density in the very inner region of $r<10^9 \cm$, which contains only $0.02M_\odot$, results from numerical limitations of not including late accretion onto the NS and from not launching late jets. We mark morphological features relevant to this study (for other aspects of this simulation, see \citealt{Braudoetal2026} on their simulation E3).}
\label{fig:SimulationInnerDensproj}
\end{center}
\end{figure}

Figure \ref{fig:SimulationInnerDensproj} zooms on the inner part of the numerical grid. The pipe and the two filaments on either side are clear. The filaments in the emission integral are projections of the pipe's envelope. The structure of a pipe between two filaments and two external arcs is qualitatively similar to the structure of the PN NGC 6720 that we present in Figure \ref{fig:NGC6720Vis}, e.g., the [Ar \textsc{iv}] image. An interesting property of the emission integral, which mimics the observed image, is that from the direction we show, the morphology is as if there were only one pair of jets. But the ejecta was powered and shaped by three pairs of jets (Table \ref{Tab:Table}). However, as \cite{Braudoetal2026} show, the Doppler shift of the two arcs is different and reveals the more complicated structure of the ejecta. We focus here only on the pipe morphology and its filaments.  
% FFFFFFFFFFFFFFFFFFFFFFFFFFFFFFFFFFFFFFFFFF
% FFFFFFFFFFFFFFFFFFFFFFFFFFFFFFFFFFFFFFFFFF

\begin{figure}[]
\begin{center}
\includegraphics[trim=0.3cm 0.4cm 0.0cm 0.0cm ,clip, angle=0, scale=1.4]{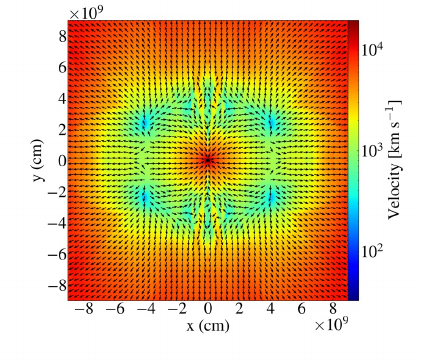} 
\caption{A velocity map in the same plane as the upper panel of Figure \ref{fig:SimulationInnerDensproj} ($z=0$) and at the same time ($t=13$). The black arrows indicate the velocity direction, and the colors present the magnitude according to the color bar. 
}
\label{fig:SimulationInnerVel}
\end{center}
\end{figure}

The velocity map that we present in Figure \ref{fig:SimulationInnerVel} shows that the velocities of the pipe and filaments are not radial. This shows that the pipe has not yet reached a homologous expansion. The inner parts are flowing back toward the center, but this might be because no additional jets were launched. The present pipe is composed of two conical pipes pointing in opposite directions, similar to those of the PNe and pre-PNe we present in Figures \ref{fig:M141} and \ref{fig:SeveralPNe}. Our expectation is that, at very late times, the inner regions of the two will merge and expand, forming a single, more cylindrical pipe extending from side to side. Later energy injection from inside, which in CCSNRs is a pulsar wind nebula and in PNe it is the fast wind from the central star, will facilitate the expansion of the narrow inner region of the pipe to form a more cylindrical-type structure.   
% FFFFFFFFFFFFFFFFFFFFFFFFFFFFFFFFFFFFFFFFFF
% FFFFFFFFFFFFFFFFFFFFFFFFFFFFFFFFFFFFFFFFFF

\begin{figure}
\centering
\includegraphics[trim=0.0cm 0.0cm 0.0cm 0.0cm ,clip, scale=0.9]{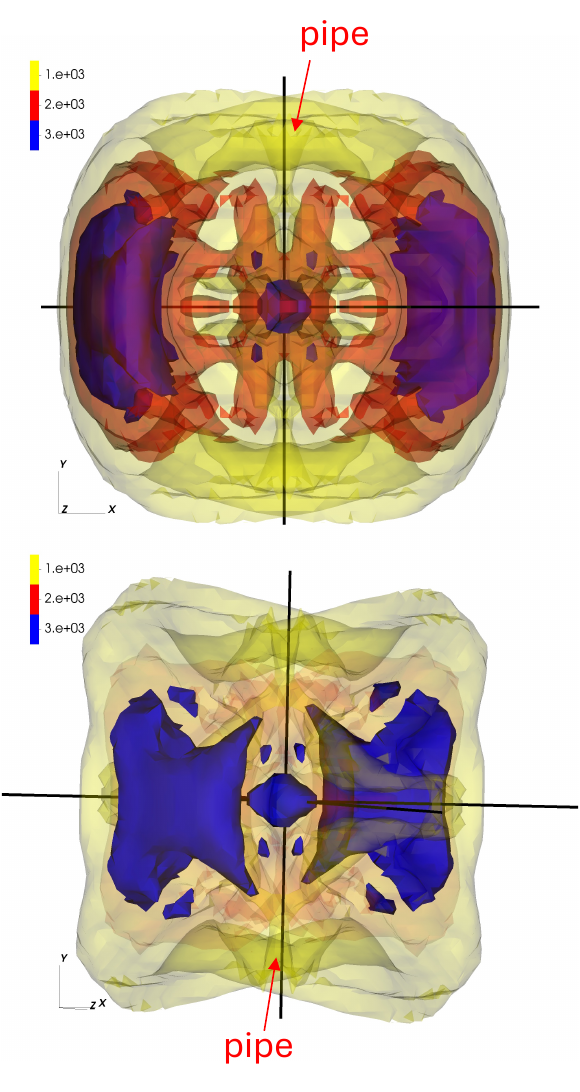} 
\caption{Three-dimensional visualizations of the gas density in the simulation presented in Figures \ref{fig:SimulationDensproj} - \ref{fig:SimulationInnerVel}, and at the same time. The images are composed of three equidensity semi-transparent surfaces, as indicated by the color bar in units of $\g \cm^{-3}$, and differ by viewing angle, as indicated by the grid axes in the lower left of each image. The black lines are the three axes of the jet pairs, two of which lie in the $y=0$ plane and appear as a single line in the upper panel. We point at the pipe.  }
\label{fig:3DVisualization}
\end{figure}

To further reveal the pipe morphology in the numerical simulation, we present a 3D visualization of the density in Figure \ref{fig:3DVisualization}. This image is composed of three equidensity, semi-transparent surfaces; the color bar shows the surface densities: yellow: $1000 \g \cm^{-3}$; red: $2000 \g \cm^{-3}$; blue: $3000 \g \cm^{-3}$. The three black lines indicate the axes of the three pairs of jets; as two axes are in the $y=0$ plane, they coincide in the upper panel of Figure \ref{fig:3DVisualization}. 
%FFFFFFFFFFFFFFFFFFFFFFFFFFFFFFFFFFF
%FFFFFFFFFFFFFFFFFFFFFFFFFFFFFFFFFFF

 We present results in a region of $\approx 10^{10} \cm$, while most CCSNRs are about a parsec and more in size, i.e., about 8 orders of magnitude difference and more. This is, of course, because of numerical constraints. However, we expect the basic morphological features to persist because the kinetic energy when we stop (for numerical reasons)  the simulation, is non-negligible. In \cite{Braudoetal2026} we showed that outside the initial radius of the star, $r \gtrsim 8 \times 10^9 \cm$, the kinetic energy of the ejecta is larger than thermal energy. This implies that the outer regions, which also expand at well above the escape speed, are already expanding in a homologous outflow and will maintain their structure until they interact with a large ambient mass; we do not study that region here. We showed in that paper that in the inner ejecta, which is what we present here, the kinetic energy is less than the thermal energy, $E_{\rm k} \lesssim E_{\rm th}$, and hence it has not yet reached a homologous expansion. Nonetheless, in \cite{Braudoetal2026} we argued that although the structure of the inner region will change quantitatively from that we present here (at $t=13 \s$), it will maintain its qualitative properties. We reasoned that, although $E_{\rm k} \lesssim E_{\rm th}$, the kinetic energy is significant and the radial momentum remains large. In \cite{Braudoetal2026}, we demonstrate that the flow maintains its main characteristics even at earlier times when the kinetic energy fraction is much lower than at the time we stop the simulation here. Specifically, we compared in that paper the inner structure at $t=2 \s$, a second after the jet activity has ceased but before the shock broke out from the star, with that at the end of the simulation here ($t=13 \s$). We found that some significant structural features persist throughout the evolution from $t=2 \s$ to $t=13 \s$, and hence we expect them to survive to the CCSNR phase. These features include the pipe. As $E_{\rm k}/E_{\rm th}$ increases with time, the ejecta will better maintain its main characteristics at later times of $t>13 \s$ than from $t=2 \s$ to $t=13 \s$. Although instabilities at early times alter small-scale structures \citep{Braudoetal2026}, the main structures, including the pipe, remain intact even at very early times.

Overall, we demonstrated here that the present CCSN simulation in the framework of the JJEM can qualitatively reproduce observed pipes. 

% ==================================
\subsection{Small scale structures}
\label{subsec:SmallScale}
% ==================================

The boundary of the pipes in the two CCSNRs is not straight. Processes that contribute to small-scale structures with random components are the vorticity and the Rayleigh-Taylor instability (RTI). The interaction of the jets with the core material excited vorticity and RTIs, with growth times of several seconds (e.g., \citealt{Braudoetal2025}). In a longer simulation than we present, these will grow to non-linear amplitudes. 

We characterize the RTI with the quantity 
\begin{equation}
f_{st} \equiv \frac{1}{\rho} \sqrt{\lvert \vec{\nabla}P \cdot \vec{\nabla}\rho \rvert} \ \text{sgn}(\vec{\nabla}P\cdot\vec{\nabla}\rho), 
\label{eq:rt}      
\end{equation}
where $\rho$ is the density, $P$ is the pressure, and ${\rm sgn} (\vec{\nabla}P\cdot\vec{\nabla}\rho)$ is the sign of the product of pressure and density gradients. 
If $f_{st}<0$, the region is unstable with a typical growth rate of $-f_{st}$, and a typical growth time of $-1/f_{st}$. 

\begin{figure}[]
\begin{center}
\includegraphics[trim=0.0cm 0.0cm 0.0cm 0.0cm ,clip, angle=0, scale=0.75]{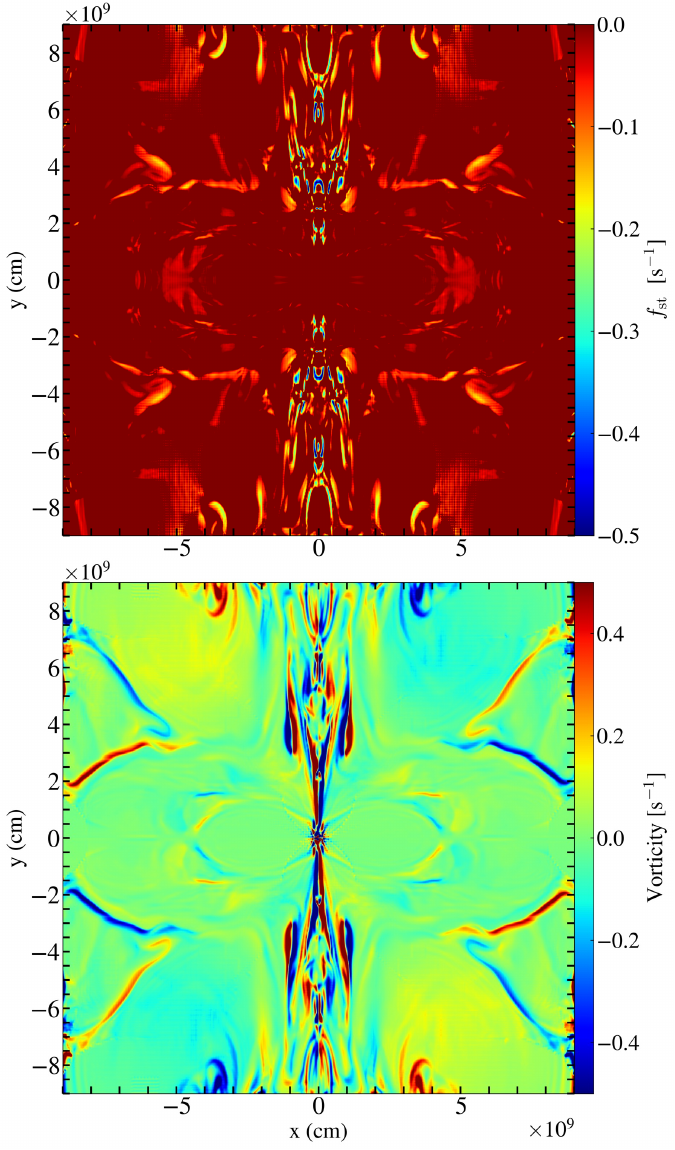} 
\caption{Rayleigh–Taylor instability (RTI) growth rate (upper panel) and vorticity (lower panel) maps at the same plane as the upper panel of Figure \ref{fig:SimulationInnerDensproj} and Figure \ref{fig:SimulationInnerVel} ($z=0$), at $t=13~\mathrm{s}$. The RTI growth rate, $f_{\rm st}$ (Equation \ref{eq:rt}), is shown in the upper panel, with values ranging from $-0.5~\mathrm{s}^{-1}$ (deep blue) to $\geq 0$ (deep red). Deep-red regions are RTI-stable, whereas blue regions are RTI-unstable, with a characteristic growth time of $-1/f_{\rm st}$. The lower panel presents the vorticity component perpendicular to the plane, $(\nabla\times\vec{v})_z$ (Equation \ref{eq:vort}). The vorticity ranges from $-0.5~\mathrm{s}^{-1}$ (deep blue; clockwise rotation) to $0.5~\mathrm{s}^{-1}$ (deep red; counter-clockwise rotation).
}
\label{fig:RTvorticitySimulation}
\end{center}
\end{figure}

In the upper panel of Figure \ref{fig:RTvorticitySimulation}, we present a map of $f_{\rm st}$ in units of $\s^{-1}$. The growth times of the regions colored blue to yellow are $\simeq 2-6 \s$, shorter than the simulation time. The highly unstable regions are along the pipe on both sides of the center. In a longer simulation, the instabilities might grow into the non-linear regime and introduce random components into the pipe's structure and its boundaries.   
% FFFFFFFFFFFFFFFFFFFFFFFFFFFFFFFFFFFFFFFFFF
% FFFFFFFFFFFFFFFFFFFFFFFFFFFFFFFFFFFFFFFFFF

In the lower panel of Figure \ref{fig:RTvorticitySimulation}, we present a map of the vorticity perpendicular to the $xy$ plane 
\begin{equation}
\omega_z=(\nabla\times\vec{v})_z ,
\label{eq:vort}      
\end{equation}
where blue is for clockwise and red is for counterclockwise rotation. There are small-scale strong vortices with overturn times of 2 to several seconds along the pipe, which are shorter than the simulation time.

The RTI and vorticity influence the small-scale structure of the pipe. The density map in the upper panel of Figure \ref{fig:SimulationInnerDensproj} shows the thin filaments and some dense clumps in the pipe and its vicinity that are shaped by the RTI and vorticity. In a future study, we will extend the simulation to the non-linear regime.

% ==================================
\section{Summary} 
\label{sec:Summary}
% ==================================

This paper is one in a series aimed at answering the question: Which is the primary explosion mechanism of CCSNe? There are two actively studied explosion mechanisms with over a dozen papers in the last year studying each (Section \ref{sec:intro}): the neutrino-driven (delayed neutrino; neutrino heating) explosion mechanism and the JJEM. At present, only the morphology of the ejecta can decisively decide between the two because the JJEM, where jets explode the star, predicts that many CCSNRs will possess prominent morphological features shaped by a pair or more of opposite jets. The neutrino-driven mechanism cannot account for most of these morphologies, particularly point-symmetric morphologies.  The identification of each new jet-pair-shaped morphological feature or a new CCSNR with a jet-pair-shaped morphology strengthens the claim that the JJEM is the primary explosion mechanism of CCSNe, including electron-capture CCSNe (\citealt{WangShishkinSoker2024}; for a recent review on electron-capture CCSNe see \citealt{WangBoetal2026RAA}). In this paper, we studied the morphological features of pipes.  

We analyzed PN and CCSN observations and one numerical simulation in four steps as follows. 
\begin{enumerate}
    \item We identified and marked the pipe morphology in two CCSNRs, The Cygnus Loop (Figure \ref{fig:CygnusLoop}) and SNR G292.0+1.8 (Figure \ref{fig:G292}). 
    \item We presented two PNe with qualitatively similar pipe morphologies to those of the CCSNRs: NGC 6720 (M 57; the Ring Nebula; Figures \ref{fig:NGC6720Vis} and \ref{fig:NGC6720IR}) and NGC 2371 (Figure \ref{fig:NGC2371}). We suggested that the highly ionized iron bar in NGC 6720, which coincides with the pipe, is gas blown by the very hot central star; this highly ionized gas fills the pipe that a pair of jets inflated. As we discussed in Section \ref{sec:PipePNe}, studies suggest that jets shaped these two PNe, possibly two or more pairs of jets.   
    \item In Section \ref{sec:formingPipes}, we presented PNe (Figures \ref{fig:M141} and \ref{fig:SeveralPNe}) where a pair of jets inflated two long and narrow lobes. We suggest that, over time, the two long lobes might merge to form a single pipe from one side to the other. We consider these PNe to support our claim that a pair of jets formed each pipe in the different objects.  
    \item In Section \ref{sec:Simulation}, we presented a 3D hydrodynamical simulation of CCSNR explosion by three pairs of equal jets in the framework of the JJEM. We demonstrated the emergence of two opposite elongated and narrow conical lobes, similar to the systems we analyzed in Section \ref{sec:formingPipes}. In a future study, we will simulate the late evolution of these lobes and their eventual merger into a pipe. 
\end{enumerate}

Going backward from step 4 to 1, we argue the following. We reproduced two opposite narrow lobes in a simulation of the JJEM (Figures \ref{fig:SimulationInnerDensproj} and \ref{fig:3DVisualization}). These lobes, which we also refer to as a pipe, are very similar to those in jet-shaped PNe (Figures \ref{fig:M141} and  \ref{fig:SeveralPNe}). It is very likely that the fully developed pipes (Figures \ref{fig:NGC6720Vis} - \ref{fig:NGC2371}) resulted from such opposite narrow lobes. Therefore, we suggest that the similar pipes in CCSNRs (Figures \ref{fig:CygnusLoop} and \ref{fig:G292}) were also shaped by pairs of jets. 

Post-explosion activity, such as an energetic magnetar (e.g., \citealt{ZhuZhang2026}) and fallback accretion that some researchers invoke (e.g.,  \citealt{ZhangJetal2026RAA}) to explain extra energy, cannot form point-symmetric morphologies. Moreover, energetic magnetars imply that jets powered the explosion (e.g., \citealt{SokerGilkis2017, Kumar2025}). Post-explosion jets that the NS launches (e.g., \citealt{WenXetal2025RAA}) are naturally explained by the JJEM as late jets. 

The short summary is that the pipe morphology in some CCSNRs supports the claim that jets shaped the ejecta, and, therefore, supports the JJEM as the primary explosion mechanism of CCSNe.

% ======================================
\section*{Acknowledgements}
% ======================================

 We thank an anonymous referee for constructive comments. 
A grant from the Pazy Foundation 2026 supported this research.
NS thanks the Charles Wolfson Academic Chair at the Technion for the support.

% =================================
% =================================

% =================================
% =================================
% =================================

%%% Below is for using the bib file

% =================================
% =================================
% =================================

%\bibliographystyle{mnras}
 \bibliography{BibReference}{}

@ARTICLE{Fryxell2000,
       author = {{Fryxell}, B. and {Olson}, K. and {Ricker}, P. and {Timmes}, F.~X. and {Zingale}, M. and {Lamb}, D.~Q. and {MacNeice}, P. and {Rosner}, R. and {Truran}, J.~W. and {Tufo}, H.},
        title = "{FLASH: An Adaptive Mesh Hydrodynamics Code for Modeling Astrophysical Thermonuclear Flashes}",
      journal = {\apjs},
         year = 2000,
        month = nov,
       volume = {131},
       number = {1},
        pages = {273-334},
          doi = {10.1086/317361},
       adsurl = {https://ui.adsabs.harvard.edu/abs/2000ApJS..131..273F}
}

@ARTICLE{Hsiaetal2014,
       author = {{Hsia}, Chih-Hao and {Chau}, Wayne and {Zhang}, Yong and {Kwok}, Sun},
        title = "{Hubble Space Telescope Observations and Geometric Models of Compact Multipolar Planetary Nebulae}",
      journal = {\apj},
         year = 2014,
        month = may,
       volume = {787},
       number = {1},
          eid = {25},
        pages = {25},
          doi = {10.1088/0004-637X/787/1/25},
       adsurl = {https://ui.adsabs.harvard.edu/abs/2014ApJ...787...25H}
}

@ARTICLE{PapishSoker2014Planar,
       author = {{Papish}, Oded and {Soker}, Noam},
        title = "{A planar jittering-jets pattern in core collapse supernova explosions}",
      journal = {\mnras},
         year = 2014,
        month = sep,
       volume = {443},
       number = {1},
        pages = {664-670},
          doi = {10.1093/mnras/stu1129},
archivePrefix = {arXiv},
       eprint = {1403.5152},
 primaryClass = {astro-ph.SR},
       adsurl = {https://ui.adsabs.harvard.edu/abs/2014MNRAS.443..664P}
}

@ARTICLE{Soker2022SNR0540,
       author = {{Soker}, Noam},
        title = "{Imprints of the Jittering Jets Explosion Mechanism in the Morphology of the Supernova Remnant SNR 0540-69.3}",
      journal = {Research in Astronomy and Astrophysics},
         year = 2022,
        month = mar,
       volume = {22},
       number = {3},
          eid = {035019},
        pages = {035019},
          doi = {10.1088/1674-4527/ac49e6},
archivePrefix = {arXiv},
       eprint = {2109.10230},
 primaryClass = {astro-ph.HE},
       adsurl = {https://ui.adsabs.harvard.edu/abs/2022RAA....22c5019S}
}

@ARTICLE{Soker2022Rev,
       author = {{Soker}, Noam},
        title = "{The Role of Jets in Exploding Supernovae and in Shaping their Remnants}",
      journal = {Research in Astronomy and Astrophysics},
         year = 2022,
        month = dec,
       volume = {22},
       number = {12},
          eid = {122003},
        pages = {122003},
          doi = {10.1088/1674-4527/ac9782},
archivePrefix = {arXiv},
       eprint = {2208.04875},
 primaryClass = {astro-ph.HE},
       adsurl = {https://ui.adsabs.harvard.edu/abs/2022RAA....22l2003S}
}

@ARTICLE{WangShishkinSoker2024,
       author = {{Wang}, Nikki Yat Ning and {Shishkin}, Dmitry and {Soker}, Noam},
        title = "{The Jittering Jets Explosion Mechanism in Electron Capture Supernovae}",
      journal = {\apj},
         year = 2024,
        month = jul,
       volume = {969},
       number = {2},
          eid = {163},
        pages = {163},
          doi = {10.3847/1538-4357/ad487f},
archivePrefix = {arXiv},
       eprint = {2401.06652},
 primaryClass = {astro-ph.HE},
       adsurl = {https://ui.adsabs.harvard.edu/abs/2024ApJ...969..163W}
}

@ARTICLE{Soker2024PNSN,
       author = {{Soker}, Noam},
        title = "{Planetary Nebula Morphologies Indicate a Jet-Driven Explosion of SN 1987A and Other Core-Collapse Supernovae}",
      journal = {Galaxies},
         year = 2024,
        month = jun,
       volume = {12},
       number = {3},
          eid = {29},
        pages = {29},
          doi = {10.3390/galaxies12030029},
archivePrefix = {arXiv},
       eprint = {2404.14843},
 primaryClass = {astro-ph.HE},
       adsurl = {https://ui.adsabs.harvard.edu/abs/2024Galax..12...29S}
}

@ARTICLE{Soker2024CF,
       author = {{Soker}, Noam},
        title = "{Comparing jet-shaped point symmetry in cluster cooling flows and supernovae}",
      journal = {The Open Journal of Astrophysics},
         year = 2024,
        month = jun,
       volume = {7},
          eid = {49},
        pages = {49},
          doi = {10.33232/001c.120279},
archivePrefix = {arXiv},
       eprint = {2403.08544},
 primaryClass = {astro-ph.HE},
       adsurl = {https://ui.adsabs.harvard.edu/abs/2024OJAp....7...49S}
}

@ARTICLE{GrichenerSoker2017,
       author = {{Grichener}, Aldana and {Soker}, Noam},
        title = "{Core collapse supernova remnants with ears}",
      journal = {\mnras},
         year = 2017,
        month = jun,
       volume = {468},
       number = {1},
        pages = {1226-1235},
          doi = {10.1093/mnras/stx534},
archivePrefix = {arXiv},
       eprint = {1610.09647},
 primaryClass = {astro-ph.HE},
       adsurl = {https://ui.adsabs.harvard.edu/abs/2017MNRAS.468.1226G}
}

@ARTICLE{Soker2024Rev,
       author = {{Soker}, Noam},
        title = "{Supernovae in 2023 (review): possible breakthroughs by late observations}",
      journal = {The Open Journal of Astrophysics},
         year = 2024,
        month = apr,
       volume = {7},
          eid = {31},
        pages = {31},
          doi = {10.33232/001c.117147},
archivePrefix = {arXiv},
       eprint = {2311.17732},
 primaryClass = {astro-ph.HE},
       adsurl = {https://ui.adsabs.harvard.edu/abs/2024OJAp....7E..31S}
}

@ARTICLE{Raymondetal2023,
       author = {{Raymond}, John C. and {Seok}, Ji Yeon and {Koo}, Bon-Chul and {Chilingarian}, Igor V. and {Grishin}, Kirill and {Caldwell}, Nelson and {Xie}, Min and {Fesen}, Robert},
        title = "{The Cygnus Loop: Shock Precursors and Electron-Ion Equilibration}",
      journal = {\apj},
         year = 2023,
        month = sep,
       volume = {954},
       number = {1},
          eid = {34},
        pages = {34},
          doi = {10.3847/1538-4357/ace692},
       adsurl = {https://ui.adsabs.harvard.edu/abs/2023ApJ...954...34R}
}

@ARTICLE{Bearetal2017,
       author = {{Bear}, Ealeal and {Grichener}, Aldana and {Soker}, Noam},
        title = "{The imprints of the last jets in core collapse supernovae}",
      journal = {\mnras},
         year = 2017,
        month = dec,
       volume = {472},
       number = {2},
        pages = {1770-1777},
          doi = {10.1093/mnras/stx2125},
archivePrefix = {arXiv},
       eprint = {1706.00003},
 primaryClass = {astro-ph.HE},
       adsurl = {https://ui.adsabs.harvard.edu/abs/2017MNRAS.472.1770B}
}

@ARTICLE{BearSoker2017,
       author = {{Bear}, Ealeal and {Soker}, Noam},
        title = "{What planetary nebulae can tell us about jets in core collapse supernovae}",
      journal = {\mnras},
         year = 2017,
        month = jun,
       volume = {468},
       number = {1},
        pages = {140-146},
          doi = {10.1093/mnras/stx431},
archivePrefix = {arXiv},
       eprint = {1611.07327},
 primaryClass = {astro-ph.SR},
       adsurl = {https://ui.adsabs.harvard.edu/abs/2017MNRAS.468..140B}
}

@ARTICLE{BearSoker2018,
       author = {{Bear}, Ealeal and {Soker}, Noam},
        title = "{Explaining the morphology of supernova remnant (SNR) 1987A with the jittering jets explosion mechanism}",
      journal = {\mnras},
         year = 2018,
        month = jul,
       volume = {478},
       number = {1},
        pages = {682-691},
          doi = {10.1093/mnras/sty1053},
archivePrefix = {arXiv},
       eprint = {1803.03946},
 primaryClass = {astro-ph.HE},
       adsurl = {https://ui.adsabs.harvard.edu/abs/2018MNRAS.478..682B}
}

@ARTICLE{ShishkinKayeSoker2024,
       author = {{Shishkin}, Dmitry and {Kaye}, Roy and {Soker}, Noam},
        title = "{Identifying Jittering Jet-shaped Ejecta in the Cygnus Loop Supernova Remnant}",
      journal = {\apj},
         year = 2024,
        month = nov,
       volume = {975},
       number = {2},
          eid = {281},
        pages = {281},
          doi = {10.3847/1538-4357/ad8138},
archivePrefix = {arXiv},
       eprint = {2408.11014},
 primaryClass = {astro-ph.HE},
       adsurl = {https://ui.adsabs.harvard.edu/abs/2024ApJ...975..281S}
}

@ARTICLE{SokerGilkis2017,
       author = {{Soker}, Noam and {Gilkis}, Avishai},
        title = "{Magnetar-powered Superluminous Supernovae Must First Be Exploded by Jets}",
      journal = {\apj},
         year = 2017,
        month = dec,
       volume = {851},
       number = {2},
          eid = {95},
        pages = {95},
          doi = {10.3847/1538-4357/aa9c83},
archivePrefix = {arXiv},
       eprint = {1708.08356},
 primaryClass = {astro-ph.HE},
       adsurl = {https://ui.adsabs.harvard.edu/abs/2017ApJ...851...95S}
}

@ARTICLE{Kumar2025,
       author = {{Kumar}, Amit},
        title = "{Insights from modelling magnetar-driven light curves of stripped-envelope supernovae}",
      journal = {\na},
         year = 2025,
        month = may,
       volume = {116},
          eid = {102346},
        pages = {102346},
          doi = {10.1016/j.newast.2024.102346},
archivePrefix = {arXiv},
       eprint = {2412.09357},
 primaryClass = {astro-ph.HE},
       adsurl = {https://ui.adsabs.harvard.edu/abs/2025NewA..11602346K}
}

@ARTICLE{Soker2025Learning,
       author = {{Soker}, Noam},
        title = "{Learning from core-collapse supernova remnants on the explosion mechanism}",
      journal = {\na},
         year = 2025,
        month = dec,
       volume = {121},
          eid = {102453},
        pages = {102453},
          doi = {10.1016/j.newast.2025.102453},
archivePrefix = {arXiv},
       eprint = {2409.13657},
 primaryClass = {astro-ph.HE},
       adsurl = {https://ui.adsabs.harvard.edu/abs/2025NewA..12102453S}
}

@ARTICLE{Soker2024UnivReview,
       author = {{Soker}, Noam},
        title = "{The Two Alternative Explosion Mechanisms of Core-Collapse Supernovae: 2024 Status Report}",
      journal = {Universe},
         year = 2024,
        month = dec,
       volume = {10},
       number = {12},
          eid = {458},
        pages = {458},
          doi = {10.3390/universe10120458},
archivePrefix = {arXiv},
       eprint = {2411.08555},
 primaryClass = {astro-ph.HE},
       adsurl = {https://ui.adsabs.harvard.edu/abs/2024Univ...10..458S}
}

@ARTICLE{Bearetal2025Puppis,
       author = {{Bear}, Ealeal and {Shishkin}, Dmitry and {Soker}, Noam},
        title = "{The Puppis A Supernova Remnant: An Early Jet-driven Neutron Star Kick followed by Jittering Jets}",
      journal = {Research in Astronomy and Astrophysics},
         year = 2025,
        month = apr,
       volume = {25},
       number = {4},
          eid = {045008},
        pages = {045008},
          doi = {10.1088/1674-4527/adc24e},
archivePrefix = {arXiv},
       eprint = {2409.11453},
 primaryClass = {astro-ph.HE},
       adsurl = {https://ui.adsabs.harvard.edu/abs/2025RAA....25d5008B}
}

@ARTICLE{Soker2025G0901,
       author = {{Soker}, Noam},
        title = "{The Morphology of Supernova Remnant G0.9+0.1 Implies Explosion by Jittering-jets}",
      journal = {Research in Astronomy and Astrophysics},
         year = 2025,
        month = nov,
       volume = {25},
       number = {11},
          eid = {115005},
        pages = {115005},
          doi = {10.1088/1674-4527/adfd23},
archivePrefix = {arXiv},
       eprint = {2504.11384},
 primaryClass = {astro-ph.HE},
       adsurl = {https://ui.adsabs.harvard.edu/abs/2025RAA....25k5005S}
}

@ARTICLE{Janka2025,
       author = {{Janka}, Hans-Thomas},
        title = "{Long-Term Multidimensional Models of Core-Collapse Supernovae: Progress and Challenges}",
      journal = {Annual Review of Nuclear and Particle Science},
         year = 2025,
        month = sep,
       volume = {75},
       number = {1},
        pages = {425-461},
          doi = {10.1146/annurev-nucl-121423-100945},
archivePrefix = {arXiv},
       eprint = {2502.14836},
 primaryClass = {astro-ph.HE},
       adsurl = {https://ui.adsabs.harvard.edu/abs/2025ARNPS..75..425J}
}

@ARTICLE{Sahaietal2007,
       author = {{Sahai}, Raghvendra and {Morris}, Mark and {S{\'a}nchez Contreras}, Carmen and {Claussen}, Mark},
        title = "{Preplanetary Nebulae: A Hubble Space Telescope Imaging Survey and a New Morphological Classification System}",
      journal = {\aj},
         year = 2007,
        month = dec,
       volume = {134},
       number = {6},
        pages = {2200-2225},
          doi = {10.1086/522944},
archivePrefix = {arXiv},
       eprint = {0707.4662},
 primaryClass = {astro-ph},
       adsurl = {https://ui.adsabs.harvard.edu/abs/2007AJ....134.2200S}
}

@ARTICLE{Braudoetal2025,
       author = {{Braudo}, Jessica and {Michaelis}, Amir and {Akashi}, Muhammad and {Soker}, Noam},
        title = "{Simulating the Shaping of Point-symmetric Structures in the Jittering Jets Explosion Mechanism}",
      journal = {\pasp},
         year = 2025,
        month = may,
       volume = {137},
       number = {5},
          eid = {054201},
        pages = {054201},
          doi = {10.1088/1538-3873/add08e},
archivePrefix = {arXiv},
       eprint = {2503.10326},
 primaryClass = {astro-ph.HE},
       adsurl = {https://ui.adsabs.harvard.edu/abs/2025PASP..137e4201B}
}

@ARTICLE{Shishkinetal2025S147,
       author = {{Shishkin}, Dmitry and {Bear}, Ealeal and {Soker}, Noam},
        title = "{Natal Kick by Early-asymmetrical Pairs of Jets to the Neutron Star of Supernova Remnant S147}",
      journal = {\apj},
         year = 2025,
        month = oct,
       volume = {992},
       number = {2},
          eid = {190},
        pages = {190},
          doi = {10.3847/1538-4357/ae0332},
archivePrefix = {arXiv},
       eprint = {2506.21548},
 primaryClass = {astro-ph.HE},
       adsurl = {https://ui.adsabs.harvard.edu/abs/2025ApJ...992..190S}
}

@ARTICLE{SokerShishkinW49B,
       author = {{Soker}, Noam and {Shishkin}, Dmitry},
        title = "{The main jet axis of the W49B supernova remnant}",
      journal = {\pasa},
         year = 2025,
        month = apr,
       volume = {42},
          eid = {e048},
        pages = {e048},
          doi = {10.1017/pasa.2025.39},
archivePrefix = {arXiv},
       eprint = {2502.09543},
 primaryClass = {astro-ph.HE},
       adsurl = {https://ui.adsabs.harvard.edu/abs/2025PASA...42...48S}
}

@ARTICLE{Akashietal2025,
       author = {{Akashi}, Muhammad and {Bear}, Ealeal and {Soker}, Noam},
        title = "{Jet-Driven Formation of Bipolar Rings in Planetary Nebulae: Numerical Simulations Inspired by NGC 1514}",
      journal = {The Open Journal of Astrophysics},
         year = 2025,
        month = sep,
       volume = {8},
          eid = {137},
        pages = {137},
          doi = {10.33232/001c.144674},
archivePrefix = {arXiv},
       eprint = {2507.23670},
 primaryClass = {astro-ph.SR},
       adsurl = {https://ui.adsabs.harvard.edu/abs/2025OJAp....8E.137A}
}

@ARTICLE{SahaiTrauger1998,
       author = {{Sahai}, Raghvendra and {Trauger}, John T.},
        title = "{Multipolar Bubbles and Jets in Low-Excitation Planetary Nebulae: Toward a New Understanding of the Formation and Shaping of Planetary Nebulae}",
      journal = {\aj},
         year = 1998,
        month = sep,
       volume = {116},
       number = {3},
        pages = {1357-1366},
          doi = {10.1086/300504},
       adsurl = {https://ui.adsabs.harvard.edu/abs/1998AJ....116.1357S}
}

@ARTICLE{Parkeretal2006,
       author = {{Parker}, Quentin A. and {Acker}, A. and {Frew}, D.~J. and {Hartley}, M. and {Peyaud}, A.~E.~J. and {Ochsenbein}, F. and {Phillipps}, S. and {Russeil}, D. and {Beaulieu}, S.~F. and {Cohen}, M. and {K{\"o}ppen}, J. and {Miszalski}, B. and {Morgan}, D.~H. and {Morris}, R.~A.~H. and {Pierce}, M.~J. and {Vaughan}, A.~E.},
        title = "{The Macquarie/AAO/Strasbourg H{\ensuremath{\alpha}} Planetary Nebula Catalogue: MASH}",
      journal = {\mnras},
         year = 2006,
        month = nov,
       volume = {373},
       number = {1},
        pages = {79-94},
          doi = {10.1111/j.1365-2966.2006.10950.x},
       adsurl = {https://ui.adsabs.harvard.edu/abs/2006MNRAS.373...79P}
}

@ARTICLE{Kwok2024Galax,
       author = {{Kwok}, Sun},
        title = "{Planetary Nebulae Research: Past, Present, and Future}",
      journal = {Galaxies},
         year = 2024,
        month = jul,
       volume = {12},
       number = {4},
          eid = {39},
        pages = {39},
          doi = {10.3390/galaxies12040039},
archivePrefix = {arXiv},
       eprint = {2408.06448},
 primaryClass = {astro-ph.SR},
       adsurl = {https://ui.adsabs.harvard.edu/abs/2024Galax..12...39K}
}

@ARTICLE{Balick1987,
       author = {{Balick}, Bruce},
        title = "{The Evolution of Planetary Nebulae. I. Structures, Ionizations, and Morphological Sequences}",
      journal = {\aj},
         year = 1987,
        month = sep,
       volume = {94},
        pages = {671},
          doi = {10.1086/114504},
       adsurl = {https://ui.adsabs.harvard.edu/abs/1987AJ.....94..671B}
}

@ARTICLE{GarciaSEguraetal2022,
       author = {{Garc{\'\i}a-Segura}, Guillermo and {Taam}, Ronald E. and {Ricker}, Paul M.},
        title = "{Common-envelope shaping of planetary nebulae - IV. From protoplanetary to planetary nebula}",
      journal = {\mnras},
         year = 2022,
        month = dec,
       volume = {517},
       number = {3},
        pages = {3822-3831},
          doi = {10.1093/mnras/stac2824},
archivePrefix = {arXiv},
       eprint = {2209.15081},
 primaryClass = {astro-ph.SR},
       adsurl = {https://ui.adsabs.harvard.edu/abs/2022MNRAS.517.3822G}
}

@ARTICLE{GarciaSEguraetal2021,
       author = {{Garc{\'\i}a-Segura}, Guillermo and {Taam}, Ronald E. and {Ricker}, Paul M.},
        title = "{Common Envelope Shaping of Planetary Nebulae. III. The Launching of Jets in Proto-Planetary Nebulae}",
      journal = {\apj},
         year = 2021,
        month = jun,
       volume = {914},
       number = {2},
          eid = {111},
        pages = {111},
          doi = {10.3847/1538-4357/abfc4e},
archivePrefix = {arXiv},
       eprint = {2104.12831},
 primaryClass = {astro-ph.SR},
       adsurl = {https://ui.adsabs.harvard.edu/abs/2021ApJ...914..111G}
}

@ARTICLE{GarciaSEguraetal2025,
       author = {{Garc{\'\i}a-Segura}, Guillermo and {Manchado}, Arturo and {Toal{\'a}}, Jes{\'u}s A. and {Guerrero}, Mart{\'\i}n A. and {Castro-Tirado}, Alberto J.},
        title = "{Planetary nebula evolution for single stellar models. The formation of neutral spikes}",
      journal = {\mnras},
         year = 2025,
        month = nov,
       volume = {543},
       number = {4},
        pages = {3867-3884},
          doi = {10.1093/mnras/staf1744},
archivePrefix = {arXiv},
       eprint = {2510.08721},
 primaryClass = {astro-ph.SR},
       adsurl = {https://ui.adsabs.harvard.edu/abs/2025MNRAS.543.3867G}
}

@ARTICLE{ShiranSoker2026,
       author = {{Shiran}, Kobi and {Soker}, Noam},
        title = "{Double shell structure in supernova 2024ggi}",
      journal = {arXiv e-prints},
         year = 2026,
        month = dec,
          eid = {arXiv:2512.16554},
        pages = {arXiv:2512.16554},
archivePrefix = {arXiv},
       eprint = {2512.16554},
 primaryClass = {astro-ph.HE},
       adsurl = {https://ui.adsabs.harvard.edu/abs/2025arXiv251216554S}
}

@ARTICLE{SokerAkashi2025,
       author = {{Soker}, Noam and {Akashi}, Muhammad},
        title = "{The explosion jets of the core-collapse supernova remnant Circinus X-1}",
      journal = {The Open Journal of Astrophysics},
         year = 2025,
        month = dec,
       volume = {8},
        pages = {54770},
          doi = {10.33232/001c.154770},
archivePrefix = {arXiv},
       eprint = {2508.10843},
 primaryClass = {astro-ph.HE},
       adsurl = {https://ui.adsabs.harvard.edu/abs/2025OJAp....854770S}
}

@ARTICLE{ShishkinMichaelis2026,
       author = {{Shishkin}, Dmitry and {Michaelis}, Amir},
        title = "{Quantifying Symmetry: Transformation Information for Planetary Nebulae and Supernova Remnants}",
      journal = {arXiv e-prints},
         year = 2026,
        month = jan,
          eid = {arXiv:2601.07913},
        pages = {arXiv:2601.07913},
          doi = {10.48550/arXiv.2601.07913},
archivePrefix = {arXiv},
       eprint = {2601.07913},
 primaryClass = {astro-ph.IM},
       adsurl = {https://ui.adsabs.harvard.edu/abs/2026arXiv260107913S}
}

@ARTICLE{Soker2026SNRJ0450,
       author = {{Soker}, Noam},
        title = "{The supernova remnant J0450.4-7050 possesses a jets-shaped point-symmetric morphology}",
      journal = {arXiv e-prints},
         year = 2026,
        month = feb,
          eid = {arXiv:2602.09993},
        pages = {arXiv:2602.09993},
          doi = {10.48550/arXiv.2602.09993},
archivePrefix = {arXiv},
       eprint = {2602.09993},
 primaryClass = {astro-ph.HE},
       adsurl = {https://ui.adsabs.harvard.edu/abs/2026arXiv260209993S}
}

@ARTICLE{Liebendorferetal2005,
       author = {{Liebend{\"o}rfer}, M. and {Rampp}, M. and {Janka}, H. -Th. and {Mezzacappa}, A.},
        title = "{Supernova Simulations with Boltzmann Neutrino Transport: A Comparison of Methods}",
      journal = {\apj},
         year = 2005,
        month = feb,
       volume = {620},
       number = {2},
        pages = {840-860},
          doi = {10.1086/427203},
archivePrefix = {arXiv},
       eprint = {astro-ph/0310662},
 primaryClass = {astro-ph},
       adsurl = {https://ui.adsabs.harvard.edu/abs/2005ApJ...620..840L}
}

@ARTICLE{PanLi2026,
       author = {{Pan}, Kuo-Chuan and {Li}, Yi-Fang},
        title = "{From Jets to Failed Supernovae: Morphologies and Gravitational-Wave Signatures in Two-Dimensional Magnetorotational Core-Collapse Supernovae}",
      journal = {arXiv e-prints},
         year = 2026,
        month = mar,
          eid = {arXiv:2603.25846},
        pages = {arXiv:2603.25846},
archivePrefix = {arXiv},
       eprint = {2603.25846},
 primaryClass = {astro-ph.HE},
       adsurl = {https://ui.adsabs.harvard.edu/abs/2026arXiv260325846P}
}

@ARTICLE{Derlopaetal2024,
       author = {{Derlopa}, S. and {Akras}, S. and {Amram}, P. and {Boumis}, P. and {Chiotellis}, A. and {de Oliveira}, C. Mendes},
        title = "{Planetary Nebula NGC 2818: Revealing its complex 3D morphology}",
      journal = {\mnras},
         year = 2024,
        month = may,
       volume = {530},
       number = {3},
        pages = {3327-3341},
          doi = {10.1093/mnras/stae1013},
archivePrefix = {arXiv},
       eprint = {2405.00169},
 primaryClass = {astro-ph.SR},
       adsurl = {https://ui.adsabs.harvard.edu/abs/2024MNRAS.530.3327D}
}

@ARTICLE{AkashiSoker2018,
       author = {{Akashi}, Muhammad and {Soker}, Noam},
        title = "{The formation of `columns crowns' by jets interacting with a circumstellar dense shell}",
      journal = {\mnras},
         year = 2018,
        month = dec,
       volume = {481},
       number = {2},
        pages = {2754-2765},
          doi = {10.1093/mnras/sty2479},
archivePrefix = {arXiv},
       eprint = {1808.00276},
 primaryClass = {astro-ph.SR},
       adsurl = {https://ui.adsabs.harvard.edu/abs/2018MNRAS.481.2754A}
}

@ARTICLE{Balicketal2020,
       author = {{Balick}, Bruce and {Frank}, Adam and {Liu}, Baowei},
        title = "{Models of the Mass-ejection Histories of Pre-planetary Nebulae. IV. Magnetized Winds and the Origins of Jets, Bullets, and FLIERs}",
      journal = {\apj},
         year = 2020,
        month = jan,
       volume = {889},
       number = {1},
          eid = {13},
        pages = {13},
          doi = {10.3847/1538-4357/ab5651},
archivePrefix = {arXiv},
       eprint = {1911.12812},
 primaryClass = {astro-ph.SR},
       adsurl = {https://ui.adsabs.harvard.edu/abs/2020ApJ...889...13B}
}

@ARTICLE{RechyGarciaetal2020,
       author = {{Rechy-Garc{\'\i}a}, J.~S. and {Guerrero}, M.~A. and {Duarte Puertas}, S. and {Chu}, Y. -H. and {Toal{\'a}}, J.~A. and {Miranda}, L.~F.},
        title = "{Kinematical investigation of possible fast collimated outflows in twelve planetary nebulae}",
      journal = {\mnras},
         year = 2020,
        month = feb,
       volume = {492},
       number = {2},
        pages = {1957-1969},
          doi = {10.1093/mnras/stz3326},
archivePrefix = {arXiv},
       eprint = {1911.11325},
 primaryClass = {astro-ph.SR},
       adsurl = {https://ui.adsabs.harvard.edu/abs/2020MNRAS.492.1957R}
}

@ARTICLE{Soker1990AJ,
       author = {{Soker}, Noam},
        title = "{On the Formation of Ansae in Planetary Nebulae}",
      journal = {\aj},
         year = 1990,
        month = jun,
       volume = {99},
        pages = {1869},
          doi = {10.1086/115465},
       adsurl = {https://ui.adsabs.harvard.edu/abs/1990AJ.....99.1869S}
}

@ARTICLE{Tafoyaetal2019,
       author = {{Tafoya}, D. and {Orosz}, G. and {Vlemmings}, W.~H.~T. and {Sahai}, R. and {P{\'e}rez-S{\'a}nchez}, A.~F.},
        title = "{Spatio-kinematical model of the collimated molecular outflow in the water-fountain nebula IRAS 16342-3814}",
      journal = {\aap},
         year = 2019,
        month = sep,
       volume = {629},
          eid = {A8},
        pages = {A8},
          doi = {10.1051/0004-6361/201834632},
archivePrefix = {arXiv},
       eprint = {1906.06328},
 primaryClass = {astro-ph.SR},
       adsurl = {https://ui.adsabs.harvard.edu/abs/2019A&A...629A...8T}
}

@ARTICLE{Morris1987,
       author = {{Morris}, Mark},
        title = "{Mechanisms for mass loss from cool stars.}",
      journal = {\pasp},
         year = 1987,
        month = nov,
       volume = {99},
        pages = {1115-1122},
          doi = {10.1086/132089},
       adsurl = {https://ui.adsabs.harvard.edu/abs/1987PASP...99.1115M}
}

@ARTICLE{MoragaBaezetal2023,
       author = {{Moraga Baez}, Paula and {Kastner}, Joel H. and {Balick}, Bruce and {Montez}, Rodolfo and {Bublitz}, Jesse},
        title = "{Panchromatic HST/WFC3 Imaging Studies of Young, Rapidly Evolving Planetary Nebulae. II. NGC 7027}",
      journal = {\apj},
         year = 2023,
        month = jan,
       volume = {942},
       number = {1},
          eid = {15},
        pages = {15},
          doi = {10.3847/1538-4357/aca401},
archivePrefix = {arXiv},
       eprint = {2210.01859},
 primaryClass = {astro-ph.GA},
       adsurl = {https://ui.adsabs.harvard.edu/abs/2023ApJ...942...15M}
}

@ARTICLE{Jones2020Galax,
       author = {{Jones}, David},
        title = "{Binary Central Stars of Planetary Nebulae}",
      journal = {Galaxies},
         year = 2020,
        month = apr,
       volume = {8},
       number = {2},
          eid = {28},
        pages = {28},
          doi = {10.3390/galaxies8020028},
       adsurl = {https://ui.adsabs.harvard.edu/abs/2020Galax...8...28J}
}

@ARTICLE{EstrellaTrujilloetal2019,
       author = {{Estrella-Trujillo}, D. and {Hern{\'a}ndez-Mart{\'\i}nez}, L. and {Vel{\'a}zquez}, P.~F. and {Esquivel}, A. and {Raga}, A.~C.},
        title = "{Hydrodynamical Models of Protoplanetary Nebulae Including the Photoionization of the Central Star}",
      journal = {\apj},
         year = 2019,
        month = may,
       volume = {876},
       number = {1},
          eid = {29},
        pages = {29},
          doi = {10.3847/1538-4357/ab12e1},
       adsurl = {https://ui.adsabs.harvard.edu/abs/2019ApJ...876...29E}
}

@ARTICLE{GarciaSeguraetal2020,
       author = {{Garc{\'\i}a-Segura}, Guillermo and {Taam}, Ronald E. and {Ricker}, Paul M.},
        title = "{Common Envelope Shaping of Planetary Nebulae. II. Magnetic Solutions and Self-collimated Outflows}",
      journal = {\apj},
         year = 2020,
        month = apr,
       volume = {893},
       number = {2},
          eid = {150},
        pages = {150},
          doi = {10.3847/1538-4357/ab8006},
archivePrefix = {arXiv},
       eprint = {2003.06073},
 primaryClass = {astro-ph.SR},
       adsurl = {https://ui.adsabs.harvard.edu/abs/2020ApJ...893..150G}
}

@ARTICLE{Clairmontetal2022,
       author = {{Clairmont}, Ryan and {Steffen}, Wolfgang and {Koning}, Nico},
        title = "{Morphokinematic modelling of the point-symmetric Cat's Eye, NGC 6543: Ring-like remnants of a precessing jet}",
      journal = {\mnras},
         year = 2022,
        month = oct,
       volume = {516},
       number = {2},
        pages = {2711-2717},
          doi = {10.1093/mnras/stac2375},
archivePrefix = {arXiv},
       eprint = {2209.01313},
 primaryClass = {astro-ph.SR},
       adsurl = {https://ui.adsabs.harvard.edu/abs/2022MNRAS.516.2711C}
}

@ARTICLE{Mirandaetal2024,
       author = {{Miranda}, Luis F. and {V{\'a}zquez}, Roberto and {Olgu{\'\i}n}, Lorenzo and {Guill{\'e}n}, Pedro F. and {Mat{\'\i}as}, Jos{\'e} M.},
        title = "{Morphokinematical study of the planetary nebula Me2-1: Unveiling its point-symmetric and unusual physical structure}",
      journal = {\aap},
         year = 2024,
        month = jul,
       volume = {687},
          eid = {A123},
        pages = {A123},
          doi = {10.1051/0004-6361/202348173},
archivePrefix = {arXiv},
       eprint = {2405.02938},
 primaryClass = {astro-ph.GA},
       adsurl = {https://ui.adsabs.harvard.edu/abs/2024A&A...687A.123M}
}

@ARTICLE{Danehkar2022,
       author = {{Danehkar}, A.},
        title = "{Morpho-kinematic Properties of Wolf-Rayet Planetary Nebulae}",
      journal = {\apjs},
         year = 2022,
        month = may,
       volume = {260},
       number = {1},
          eid = {14},
        pages = {14},
          doi = {10.3847/1538-4365/ac5cca},
archivePrefix = {arXiv},
       eprint = {2107.03994},
 primaryClass = {astro-ph.SR},
       adsurl = {https://ui.adsabs.harvard.edu/abs/2022ApJS..260...14D}
}

@ARTICLE{Sahaietal2024,
       author = {{Sahai}, Raghvendra and {Alcolea}, Javier and {Balick}, Bruce and {Blackman}, Eric G. and {Bujarrabal}, Valentin and {Castro-Carrizo}, Arancha and {De Marco}, Orsola and {Kastner}, Joel and {Kim}, Hyosun and {Lagadec}, Eric and {Lee}, Chin-Fei and {Sabin}, Laurence and {Santander-Garcia}, M. and {S{\'a}nchez Contreras}, Carmen and {Tafoya}, Daniel and {Ueta}, Toshiya and {Vlemmings}, Wouter and {Zijlstra}, Albert},
        title = "{High-Speed Outflows and Dusty Disks during the AGB to PN Transition: The PANORAMA survey}",
      journal = {arXiv e-prints},
         year = 2024,
        month = sep,
          eid = {arXiv:2409.06038},
        pages = {arXiv:2409.06038},
archivePrefix = {arXiv},
       eprint = {2409.06038},
 primaryClass = {astro-ph.SR},
       adsurl = {https://ui.adsabs.harvard.edu/abs/2024arXiv240906038S}
}

@ARTICLE{Mirandaetal1998,
       author = {{Miranda}, Luis F. and {Torrelles}, Jose M. and {Guerrero}, Martin A. and {Aaquist}, Orla B. and {Eiroa}, Carlos},
        title = "{The nature and structure of the emission line nebula K3-35: a very young planetary nebula with precessing bipolar jet-like outflows?}",
      journal = {\mnras},
         year = 1998,
        month = jul,
       volume = {298},
       number = {1},
        pages = {243-250},
          doi = {10.1046/j.1365-8711.1998.01611.x},
       adsurl = {https://ui.adsabs.harvard.edu/abs/1998MNRAS.298..243M}
}

@ARTICLE{Sahaietal2005,
       author = {{Sahai}, R. and {Le Mignant}, D. and {S{\'a}nchez Contreras}, C. and {Campbell}, R.~D. and {Chaffee}, F.~H.},
        title = "{Sculpting a Pre-planetary Nebula with a Precessing Jet: IRAS 16342-3814}",
      journal = {\apjl},
         year = 2005,
        month = mar,
       volume = {622},
       number = {1},
        pages = {L53-L56},
          doi = {10.1086/429586},
       adsurl = {https://ui.adsabs.harvard.edu/abs/2005ApJ...622L..53S}
}

@ARTICLE{Guerreroetal1998,
       author = {{Guerrero}, M.~A. and {Manchado}, A.},
        title = "{A Precessing Collimated Outflow in the Quadrupolar Planetary Nebula NGC 6881}",
      journal = {\apj},
         year = 1998,
        month = nov,
       volume = {508},
       number = {1},
        pages = {262-267},
          doi = {10.1086/306407},
       adsurl = {https://ui.adsabs.harvard.edu/abs/1998ApJ...508..262G}
}

@ARTICLE{Ablimit2024,
       author = {{Ablimit}, Iminhaji},
        title = "{Formation of neutron stars inside planetary nebulae via accretion-induced collapse and core-merger-induced collapse from white dwarf binaries}",
      journal = {arXiv e-prints},
         year = 2024,
        month = jul,
          eid = {arXiv:2407.03985},
        pages = {arXiv:2407.03985},
          doi = {10.48550/arXiv.2407.03985},
archivePrefix = {arXiv},
       eprint = {2407.03985},
 primaryClass = {astro-ph.HE},
       adsurl = {https://ui.adsabs.harvard.edu/abs/2024arXiv240703985A}
}

@ARTICLE{Sowickaetal2017,
       author = {{Sowicka}, Paulina and {Jones}, David and {Corradi}, Romano L.~M. and {Wesson}, Roger and {Garc{\'\i}a-Rojas}, Jorge and {Santander-Garc{\'\i}a}, Miguel and {Boffin}, Henri M.~J. and {Rodr{\'\i}guez-Gil}, Pablo},
        title = "{The planetary nebula IC 4776 and its post-common-envelope binary central star}",
      journal = {\mnras},
         year = 2017,
        month = nov,
       volume = {471},
       number = {3},
        pages = {3529-3546},
          doi = {10.1093/mnras/stx1697},
archivePrefix = {arXiv},
       eprint = {1706.08766},
 primaryClass = {astro-ph.SR},
       adsurl = {https://ui.adsabs.harvard.edu/abs/2017MNRAS.471.3529S}
}

@ARTICLE{RechyGarciaetal2019,
       author = {{Rechy-Garc{\'\i}a}, J.~S. and {Pe{\~n}a}, M. and {Vel{\'a}zquez}, P.~F.},
        title = "{3D hydrodynamical models of point-symmetric planetary nebulae: the special case of H 1-67}",
      journal = {\mnras},
         year = 2019,
        month = jan,
       volume = {482},
       number = {1},
        pages = {1163-1175},
          doi = {10.1093/mnras/sty2758},
archivePrefix = {arXiv},
       eprint = {1810.07271},
 primaryClass = {astro-ph.SR},
       adsurl = {https://ui.adsabs.harvard.edu/abs/2019MNRAS.482.1163R}
}

@ARTICLE{Guerreoetal2021,
       author = {{Guerrero}, M.~A. and {Cazzoli}, S. and {Rechy-Garc{\'\i}a}, J.~S. and {Ramos-Larios}, G. and {Montoro-Molina}, B. and {G{\'o}mez-Gonz{\'a}lez}, V.~M.~A. and {Toal{\'a}}, J.~A. and {Fang}, X.},
        title = "{Tomography of the Unique Ongoing Jet in the Planetary Nebula NGC 2392}",
      journal = {\apj},
         year = 2021,
        month = mar,
       volume = {909},
       number = {1},
          eid = {44},
        pages = {44},
          doi = {10.3847/1538-4357/abe2aa},
archivePrefix = {arXiv},
       eprint = {2102.01093},
 primaryClass = {astro-ph.SR},
       adsurl = {https://ui.adsabs.harvard.edu/abs/2021ApJ...909...44G}
}

@ARTICLE{Akashietal2018,
       author = {{Akashi}, Muhammad and {Bear}, Ealeal and {Soker}, Noam},
        title = "{Forming H-shaped and barrel-shaped nebulae with interacting jets}",
      journal = {\mnras},
         year = 2018,
        month = apr,
       volume = {475},
       number = {4},
        pages = {4794-4808},
          doi = {10.1093/mnras/sty029},
archivePrefix = {arXiv},
       eprint = {1712.07156},
 primaryClass = {astro-ph.SR},
       adsurl = {https://ui.adsabs.harvard.edu/abs/2018MNRAS.475.4794A}
}

@ARTICLE{Boffinetal2012,
       author = {{Boffin}, Henri M.~J. and {Miszalski}, Brent and {Rauch}, Thomas and {Jones}, David and {Corradi}, Romano L.~M. and {Napiwotzki}, Ralf and {Day-Jones}, Avril C. and {K{\"o}ppen}, Joachim},
        title = "{An Interacting Binary System Powers Precessing Outflows of an Evolved Star}",
      journal = {Science},
         year = 2012,
        month = nov,
       volume = {338},
       number = {6108},
        pages = {773},
          doi = {10.1126/science.1225386},
archivePrefix = {arXiv},
       eprint = {1211.2200},
 primaryClass = {astro-ph.GA},
       adsurl = {https://ui.adsabs.harvard.edu/abs/2012Sci...338..773B}
}

@ARTICLE{SokerShiran2025,
       author = {{Soker}, Noam and {Shiran}, Kobi},
        title = "{Multiple Shells in Supernova 2023ixf Support the Jittering Jets Explosion Mechanism}",
      journal = {\apjl},
         year = 2025,
        month = dec,
       volume = {995},
       number = {1},
          eid = {L3},
        pages = {L3},
          doi = {10.3847/2041-8213/ae21da},
archivePrefix = {arXiv},
       eprint = {2510.18782},
 primaryClass = {astro-ph.HE},
       adsurl = {https://ui.adsabs.harvard.edu/abs/2025ApJ...995L...3S}
}

@ARTICLE{Zimmermanetal2024,
       author = {{Zimmerman}, E.~A. and {Irani}, I. and {Chen}, P. and {Gal-Yam}, A. and {Schulze}, S. and {Perley}, D.~A. and {Sollerman}, J. and {Filippenko}, A.~V. and {Shenar}, T. and {Yaron}, O. and {Shahaf}, S. and {Bruch}, R.~J. and {Ofek}, E.~O. and {De Cia}, A. and {Brink}, T.~G. and {Yang}, Y. and {Vasylyev}, S.~S. and {Ben Ami}, S. and {Aubert}, M. and {Badash}, A. and {Bloom}, J.~S. and {Brown}, P.~J. and {De}, K. and {Dimitriadis}, G. and {Fransson}, C. and {Fremling}, C. and {Hinds}, K. and {Horesh}, A. and {Johansson}, J.~P. and {Kasliwal}, M.~M. and {Kulkarni}, S.~R. and {Kushnir}, D. and {Martin}, C. and {Matuzewski}, M. and {McGurk}, R.~C. and {Miller}, A.~A. and {Morag}, J. and {Neil}, J.~D. and {Nugent}, P.~E. and {Post}, R.~S. and {Prusinski}, N.~Z. and {Qin}, Y. and {Raichoor}, A. and {Riddle}, R. and {Rowe}, M. and {Rusholme}, B. and {Sfaradi}, I. and {Sjoberg}, K.~M. and {Soumagnac}, M. and {Stein}, R.~D. and {Strotjohann}, N.~L. and {Terwel}, J.~H. and {Wasserman}, T. and {Wise}, J. and {Wold}, A. and {Yan}, L. and {Zhang}, K.},
        title = "{The complex circumstellar environment of supernova 2023ixf}",
      journal = {\nat},
         year = 2024,
        month = mar,
       volume = {627},
       number = {8005},
        pages = {759-762},
          doi = {10.1038/s41586-024-07116-6},
archivePrefix = {arXiv},
       eprint = {2310.10727},
 primaryClass = {astro-ph.HE},
       adsurl = {https://ui.adsabs.harvard.edu/abs/2024Natur.627..759Z}
}

@ARTICLE{ChenTWetal2025,
       author = {{Chen}, Ting-Wan and {Yang}, Sheng and {Srivastav}, Shubham and {Moriya}, Takashi J. and {Smartt}, Stephen J. and {Rest}, Sofia and {Rest}, Armin and {Lin}, Hsing Wen and {Miao}, Hao-Yu and {Cheng}, Yu-Chi and {Aryan}, Amar and {Cheng}, Chia-Yu and {Fraser}, Morgan and {Huang}, Li-Ching and {Lee}, Meng-Han and {Lai}, Cheng-Han and {Liu}, Yu-Hsuan and {Sankar. K}, Aiswarya and {Smith}, Ken W. and {Stevance}, Heloise F. and {Wang}, Ze-Ning and {Anderson}, Joseph P. and {Angus}, Charlotte R. and {de Boer}, Thomas and {Chambers}, Kenneth and {Duan}, Hao-Yuan and {Erasmus}, Nicolas and {Fulton}, Michael and {Gao}, Hua and {Herman}, Joanna and {Hou}, Wei-Jie and {Hsiao}, Hsiang-Yao and {Huber}, Mark E. and {Lin}, Chien-Cheng and {Lin}, Hung-Chin and {Magnier}, Eugene A. and {Man}, Ka Kit and {Moore}, Thomas and {Ngeow}, Chow-Choong and {Nicholl}, Matt and {Ou}, Po-Sheng and {Pignata}, Giuliano and {Shiau}, Yu-Chien and {Sommer}, Julian Silvester and {Tonry}, John L. and {Wang}, Xiao-Feng and {Wainscoat}, Richard and {Young}, David R. and {Yeh}, You-Ting and {Zhang}, Jujia},
        title = "{Discovery and Extensive Follow-up of SN 2024ggi, a Nearby Type IIP Supernova in NGC 3621}",
      journal = {\apj},
         year = 2025,
        month = apr,
       volume = {983},
       number = {1},
          eid = {86},
        pages = {86},
          doi = {10.3847/1538-4357/adb428},
archivePrefix = {arXiv},
       eprint = {2406.09270},
 primaryClass = {astro-ph.HE},
       adsurl = {https://ui.adsabs.harvard.edu/abs/2025ApJ...983...86C}
}

@ARTICLE{Giudicietal2026,
       author = {{Giudici}, Beatrice and {Gabler}, Michael and {Janka}, Hans-Thomas},
        title = "{Hydrodynamic instabilities in long-term three-dimensional simulations of neutrino-driven supernovae of 13 red supergiant progenitors}",
      journal = {arXiv e-prints},
         year = 2025,
        month = nov,
          eid = {arXiv:2511.11796},
        pages = {arXiv:2511.11796},
archivePrefix = {arXiv},
       eprint = {2511.11796},
 primaryClass = {astro-ph.HE},
       adsurl = {https://ui.adsabs.harvard.edu/abs/2025arXiv251111796G}
}

@ARTICLE{LuoZhaKajino2026,
       author = {{Luo}, Y. and {Zha}, S. and {Kajino}, T.},
        title = "{Systematic study of the impact of magnetic fields on neutrino transport in core-collapse supernovae}",
      journal = {\prd},
         year = 2026,
        month = jan,
       volume = {113},
       number = {2},
          eid = {023024},
        pages = {023024},
          doi = {10.1103/7ytg-wzl8},
archivePrefix = {arXiv},
       eprint = {2512.10417},
 primaryClass = {astro-ph.HE},
       adsurl = {https://ui.adsabs.harvard.edu/abs/2026PhRvD.113b3024L}
}

@ARTICLE{Rusakovetal2026,
       author = {{Rusakov}, Aleksandr and {Burrows}, Adam S. and {Wang}, Tianshu and {Vartanyan}, David},
        title = "{An Exploration of the Equation of State Dependence of Core-Collapse Supernova Explosion Outcomes and Signatures}",
      journal = {arXiv e-prints},
         year = 2026,
        month = feb,
          eid = {arXiv:2602.09025},
        pages = {arXiv:2602.09025},
archivePrefix = {arXiv},
       eprint = {2602.09025},
 primaryClass = {astro-ph.HE},
       adsurl = {https://ui.adsabs.harvard.edu/abs/2026arXiv260209025R}
}

@ARTICLE{Murphyetal2026,
       author = {{Murphy}, R. Daniel and {Brinkman}, Elle and {Richardson}, Colter J. and {Semenak}, Evan and {Mezzacappa}, Anthony and {Marronetti}, Pedro and {Lentz}, Eric J. and {Bruenn}, Stephen W.},
        title = "{Gravitational Waves as a Probe of Core Collapse Supernova Progenitor Structure}",
      journal = {arXiv e-prints},
         year = 2025,
        month = nov,
          eid = {arXiv:2511.21895},
        pages = {arXiv:2511.21895},
          doi = {10.48550/arXiv.2511.21895},
archivePrefix = {arXiv},
       eprint = {2511.21895},
 primaryClass = {astro-ph.HE},
       adsurl = {https://ui.adsabs.harvard.edu/abs/2025arXiv251121895M}
}

@ARTICLE{VarmaMuller2026,
       author = {{Varma}, Vishnu and {M{\"u}ller}, Bernhard},
        title = "{Resolution Dependence in Magnetohydrodynamic Simulations of Neutrino-Driven Core-Collapse Supernovae}",
      journal = {arXiv e-prints},
         year = 2026,
        month = jan,
          eid = {arXiv:2601.18357},
        pages = {arXiv:2601.18357},
          doi = {10.48550/arXiv.2601.18357},
archivePrefix = {arXiv},
       eprint = {2601.18357},
 primaryClass = {astro-ph.HE},
       adsurl = {https://ui.adsabs.harvard.edu/abs/2026arXiv260118357V}
}

@ARTICLE{Wessonetal2026,
       author = {{Wesson}, R. and {Gabler}, M. and {Lyons}, M. and {Wildman}, J. and {Matsuura}, Mikako and {Janka}, H.-T. and {Giudici}, B. and {Cigan}, P. and {Gomez}, H.~L. and {Indebetouw}, R. and {Richards}, A.~M.~S. and {Wongwathanarat}, A.},
        title = "{3D insights into SN 1987A: ALMA observations compared to hydrodynamical explosion simulations}",
      journal = {arXiv e-prints},
         year = 2026,
        month = feb,
          eid = {arXiv:2602.11259},
        pages = {arXiv:2602.11259},
          doi = {10.48550/arXiv.2602.11259},
archivePrefix = {arXiv},
       eprint = {2602.11259},
 primaryClass = {astro-ph.HE},
       adsurl = {https://ui.adsabs.harvard.edu/abs/2026arXiv260211259W}
}

@ARTICLE{Soker2026Failed,
       author = {{Soker}, Noam},
        title = "{The failed failed-supernova scenario of M31-2014-DS1}",
      journal = {The Open Journal of Astrophysics},
         year = 2026,
        month = jul,
       volume = {9},
        pages = {65498},
          doi = {10.33232/001c.165498},
archivePrefix = {arXiv},
       eprint = {2601.14497},
 primaryClass = {astro-ph.HE},
       adsurl = {https://ui.adsabs.harvard.edu/abs/2026OJAp....965498S}
}

@ARTICLE{Kastneretal2025,
       author = {{Kastner}, Joel H. and {Moraga Baez}, Paula and {Balick}, Bruce and {Montez}, Jr., Rodolfo and {Gieser}, Caroline and {Matsuura}, Mikako and {Nordhaus}, Jason and {Santander-Garcia}, Miguel},
        title = "{JWST/NIRCam Imaging of the Bipolar Planetary Nebula NGC 6537: The (Infra)red Spider, Revealed}",
      journal = {\apj},
         year = 2025,
        month = nov,
       volume = {993},
       number = {1},
          eid = {79},
        pages = {79},
          doi = {10.3847/1538-4357/ae0706},
archivePrefix = {arXiv},
       eprint = {2509.11042},
 primaryClass = {astro-ph.GA},
       adsurl = {https://ui.adsabs.harvard.edu/abs/2025ApJ...993...79K}
}

@ARTICLE{AkashiSoker2026a,
       author = {{Akashi}, Muhammad and {Soker}, Noam},
        title = "{Simulating the jittering-jets explosion mechanism: circum-jet rings account for observed core-collapse supernova remnant morphologies}",
      journal = {arXiv e-prints},
         year = 2026,
        month = mar,
          eid = {arXiv:2603.29527},
        pages = {arXiv:2603.29527},
archivePrefix = {arXiv},
       eprint = {2603.29527},
 primaryClass = {astro-ph.HE},
       adsurl = {https://ui.adsabs.harvard.edu/abs/2026arXiv260329527A}
}

@ARTICLE{SokerBisker2006,
       author = {{Soker}, Noam and {Bisker}, Gili},
        title = "{Bubbles in planetary nebulae and clusters of galaxies: jet bending}",
      journal = {\mnras},
         year = 2006,
        month = jul,
       volume = {369},
       number = {3},
        pages = {1115-1122},
          doi = {10.1111/j.1365-2966.2006.10313.x},
archivePrefix = {arXiv},
       eprint = {astro-ph/0601032},
 primaryClass = {astro-ph},
       adsurl = {https://ui.adsabs.harvard.edu/abs/2006MNRAS.369.1115S}
}

@ARTICLE{Soker2025RobustJetsRAA,
       author = {{Soker}, Noam},
        title = "{Jets are the Most Robust Observable Ingredient of Common Envelope Evolution}",
      journal = {Research in Astronomy and Astrophysics},
         year = 2025,
        month = feb,
       volume = {25},
       number = {2},
          eid = {025023},
        pages = {025023},
          doi = {10.1088/1674-4527/adb15b},
archivePrefix = {arXiv},
       eprint = {2412.04017},
 primaryClass = {astro-ph.SR},
       adsurl = {https://ui.adsabs.harvard.edu/abs/2025RAA....25b5023S}
}

@ARTICLE{Jones2025,
       author = {{Jones}, D.},
        title = "{Post-common-envelope planetary nebulae}",
      journal = {Contributions of the Astronomical Observatory Skalnate Pleso},
         year = 2025,
        month = apr,
       volume = {55},
       number = {3},
        pages = {200-210},
          doi = {10.31577/caosp.2025.55.3.200},
archivePrefix = {arXiv},
       eprint = {2411.06831},
 primaryClass = {astro-ph.SR},
       adsurl = {https://ui.adsabs.harvard.edu/abs/2025CoSka..55c.200J}
}

@ARTICLE{Miszalski2019ic,
       author = {{Miszalski}, B. and {Manick}, R. and {Van Winckel}, H. and {Miko{\l}ajewska}, J.},
        title = "{The post-common-envelope binary nucleus of the planetary nebula IC 4776: neither an anomalously long orbital period nor a Wolf-Rayet binary}",
      journal = {\mnras},
         year = 2019,
        month = jul,
       volume = {487},
       number = {1},
        pages = {1040-1046},
          doi = {10.1093/mnras/stz1315},
archivePrefix = {arXiv},
       eprint = {1905.03714},
 primaryClass = {astro-ph.SR},
       adsurl = {https://ui.adsabs.harvard.edu/abs/2019MNRAS.487.1040M}
}

@ARTICLE{Bhaleraoetal2019,
       author = {{Bhalerao}, Jayant and {Park}, Sangwook and {Schenck}, Andrew and {Post}, Seth and {Hughes}, John P.},
        title = "{Detailed X-Ray Mapping of the Shocked Ejecta and Circumstellar Medium in the Galactic Core-collapse Supernova Remnant G292.0+1.8}",
      journal = {\apj},
         year = 2019,
        month = feb,
       volume = {872},
       number = {1},
          eid = {31},
        pages = {31},
          doi = {10.3847/1538-4357/aafafd},
archivePrefix = {arXiv},
       eprint = {1812.09605},
 primaryClass = {astro-ph.HE},
       adsurl = {https://ui.adsabs.harvard.edu/abs/2019ApJ...872...31B}
}

@ARTICLE{Parketal2007,
       author = {{Park}, Sangwook and {Hughes}, John P. and {Slane}, Patrick O. and {Burrows}, David N. and {Gaensler}, B.~M. and {Ghavamian}, Parviz},
        title = "{A Half-Megasecond Chandra Observation of the Oxygen-rich Supernova Remnant G292.0+1.8}",
      journal = {\apjl},
         year = 2007,
        month = dec,
       volume = {670},
       number = {2},
        pages = {L121-L124},
          doi = {10.1086/524406},
archivePrefix = {arXiv},
       eprint = {0710.2902},
 primaryClass = {astro-ph},
       adsurl = {https://ui.adsabs.harvard.edu/abs/2007ApJ...670L.121P}
}

@ARTICLE{Temimetal2022,
       author = {{Temim}, Tea and {Slane}, Patrick and {Raymond}, John C. and {Patnaude}, Daniel and {Murray}, Emily and {Ghavamian}, Parviz and {Renzo}, Mathieu and {Jacovich}, Taylor},
        title = "{SNR G292.0+1.8: A Remnant of a Low-mass Progenitor Stripped-envelope Supernova}",
      journal = {\apj},
         year = 2022,
        month = jun,
       volume = {932},
       number = {1},
          eid = {26},
        pages = {26},
          doi = {10.3847/1538-4357/ac6bf4},
archivePrefix = {arXiv},
       eprint = {2205.01798},
 primaryClass = {astro-ph.HE},
       adsurl = {https://ui.adsabs.harvard.edu/abs/2022ApJ...932...26T}
}

@ARTICLE{GonzalezSafiHarb2003,
       author = {{Gonzalez}, Marjorie and {Safi-Harb}, Samar},
        title = "{New Constraints on the Energetics, Progenitor Mass, and Age of the Supernova Remnant G292.0+1.8 Containing PSR J1124-5916}",
      journal = {\apjl},
         year = 2003,
        month = feb,
       volume = {583},
       number = {2},
        pages = {L91-L94},
          doi = {10.1086/368122},
archivePrefix = {arXiv},
       eprint = {astro-ph/0301193},
 primaryClass = {astro-ph},
       adsurl = {https://ui.adsabs.harvard.edu/abs/2003ApJ...583L..91G}
}

@ARTICLE{Parketal2002,
       author = {{Park}, Sangwook and {Roming}, Peter W.~A. and {Hughes}, John P. and {Slane}, Patrick O. and {Burrows}, David N. and {Garmire}, Gordon P. and {Nousek}, John A.},
        title = "{The Structure of the Oxygen-rich Supernova Remnant G292.0+1.8 from Chandra X-Ray Images: Shocked Ejecta and Circumstellar Medium}",
      journal = {\apjl},
         year = 2002,
        month = jan,
       volume = {564},
       number = {1},
        pages = {L39-L43},
          doi = {10.1086/338861},
archivePrefix = {arXiv},
       eprint = {astro-ph/0112033},
 primaryClass = {astro-ph},
       adsurl = {https://ui.adsabs.harvard.edu/abs/2002ApJ...564L..39P}
}

@ARTICLE{Parketal2004,
       author = {{Park}, Sangwook and {Hughes}, John P. and {Slane}, Patrick O. and {Burrows}, David N. and {Roming}, Peter W.~A. and {Nousek}, John A. and {Garmire}, Gordon P.},
        title = "{Nucleosynthesis in the Oxygen-rich Supernova Remnant G292.0+1.8 from Chandra X-Ray Spectroscopy}",
      journal = {\apjl},
         year = 2004,
        month = feb,
       volume = {602},
       number = {1},
        pages = {L33-L36},
          doi = {10.1086/382276},
archivePrefix = {arXiv},
       eprint = {astro-ph/0312637},
 primaryClass = {astro-ph},
       adsurl = {https://ui.adsabs.harvard.edu/abs/2004ApJ...602L..33P}
}

@ARTICLE{Leeetal2010,
       author = {{Lee}, Jae-Joon and {Park}, Sangwook and {Hughes}, John P. and {Slane}, Patrick O. and {Gaensler}, B.~M. and {Ghavamian}, Parviz and {Burrows}, David N.},
        title = "{The Outer Shock of the Oxygen-Rich Supernova Remnant G292.0+1.8: Evidence for the Interaction with the Stellar Winds from Its Massive Progenitor}",
      journal = {\apj},
         year = 2010,
        month = mar,
       volume = {711},
       number = {2},
        pages = {861-869},
          doi = {10.1088/0004-637X/711/2/861},
archivePrefix = {arXiv},
       eprint = {1001.4045},
 primaryClass = {astro-ph.GA},
       adsurl = {https://ui.adsabs.harvard.edu/abs/2010ApJ...711..861L}
}

@ARTICLE{Leeetal2009,
       author = {{Lee}, Ho-Gyu and {Koo}, Bon-Chul and {Moon}, Dae-Sik and {Sakon}, Itsuki and {Onaka}, Takashi and {Jeong}, Woong-Seob and {Kaneda}, Hidehiro and {Nozawa}, Takaya and {Kozasa}, Takashi},
        title = "{AKARI Infrared Observations of the Supernova Remnant G292.0+1.8: Unveiling Circumstellar Medium and Supernova Ejecta}",
      journal = {\apj},
         year = 2009,
        month = nov,
       volume = {706},
       number = {1},
        pages = {441-453},
          doi = {10.1088/0004-637X/706/1/441},
archivePrefix = {arXiv},
       eprint = {0910.0922},
 primaryClass = {astro-ph.GA},
       adsurl = {https://ui.adsabs.harvard.edu/abs/2009ApJ...706..441L}
}

@ARTICLE{Ghavamianetal2005,
       author = {{Ghavamian}, Parviz and {Hughes}, John P. and {Williams}, T.~B.},
        title = "{Exploring the Kinematics of the Oxygen-rich Supernova Remnant G292.0+1.8: Ejecta Shells, Fast-moving Knots, and Shocked Circumstellar Material}",
      journal = {\apj},
         year = 2005,
        month = dec,
       volume = {635},
       number = {1},
        pages = {365-380},
          doi = {10.1086/497283},
archivePrefix = {arXiv},
       eprint = {astro-ph/0508309},
 primaryClass = {astro-ph},
       adsurl = {https://ui.adsabs.harvard.edu/abs/2005ApJ...635..365G}
}

@ARTICLE{Ghavamianetal2012,
       author = {{Ghavamian}, Parviz and {Long}, Knox S. and {Blair}, William P. and {Park}, Sangwook and {Fesen}, Robert and {Gaensler}, B.~M. and {Hughes}, John P. and {Rho}, Jeonghee and {Winkler}, P. Frank},
        title = "{Spitzer Imaging and Spectral Mapping of the Oxygen-rich Supernova Remnant G292.0+1.8}",
      journal = {\apj},
         year = 2012,
        month = may,
       volume = {750},
       number = {1},
          eid = {39},
        pages = {39},
          doi = {10.1088/0004-637X/750/1/39},
archivePrefix = {arXiv},
       eprint = {1202.5823},
 primaryClass = {astro-ph.GA},
       adsurl = {https://ui.adsabs.harvard.edu/abs/2012ApJ...750...39G}
}

@ARTICLE{LonkXietal2022,
       author = {{Long}, Xi and {Patnaude}, Daniel J. and {Plucinsky}, Paul P. and {Gaetz}, Terrance J.},
        title = "{The Proper Motion of the Pulsar J1124-5916 in the Galactic Supernova Remnant G292.0+1.8}",
      journal = {\apj},
         year = 2022,
        month = jun,
       volume = {932},
       number = {2},
          eid = {117},
        pages = {117},
          doi = {10.3847/1538-4357/ac704b},
archivePrefix = {arXiv},
       eprint = {2205.07951},
 primaryClass = {astro-ph.HE},
       adsurl = {https://ui.adsabs.harvard.edu/abs/2022ApJ...932..117L}
}

@ARTICLE{Naritaetal2024,
       author = {{Narita}, Takuto and {Uchida}, Hiroyuki and {Vink}, Jacco and {Katsuda}, Satoru and {Umeda}, Hideyuki and {Yoshida}, Takashi and {Sato}, Toshiki and {Matsunaga}, Kai and {Tsuru}, Takeshi Go},
        title = "{Evidence for Type Ib/c Origin of the Supernova Remnant G292.0+1.8}",
      journal = {\apj},
         year = 2024,
        month = nov,
       volume = {976},
       number = {1},
          eid = {146},
        pages = {146},
          doi = {10.3847/1538-4357/ad7e17},
archivePrefix = {arXiv},
       eprint = {2409.17542},
 primaryClass = {astro-ph.HE},
       adsurl = {https://ui.adsabs.harvard.edu/abs/2024ApJ...976..146N}
}

@ARTICLE{Lemiereetal2026,
       author = {{Lemi{\`e}re}, A. and {Castelletti}, G. and {Maza}, N.~L.},
        title = "{Unlocking the radio- {\ensuremath{\gamma}} spectrum of the pulsar wind nebula around PSR J1124─5916 in SNR G292.0+1.8}",
      journal = {\aap},
         year = 2026,
        month = jan,
       volume = {705},
          eid = {A218},
        pages = {A218},
          doi = {10.1051/0004-6361/202556224},
archivePrefix = {arXiv},
       eprint = {2510.23747},
 primaryClass = {astro-ph.HE},
       adsurl = {https://ui.adsabs.harvard.edu/abs/2026A&A...705A.218L}
}
  \bibliographystyle{aasjournal}

\end{document}